\documentclass[11pt,nofootinbib,showpacs,floatfix]{revtex4}
\usepackage{amsfonts,amssymb,amsmath,graphicx,multirow,dsfont}
\def\dac{\displaystyle\frac}

\def\[{\left[}
\def\]{\right]}
\def\({\left(}
\def\){\right)}

\newcommand{\diag}{\mathop{\rm diag}\nolimits}

\begin{document}

\baselineskip7mm
\title{Cosmological dynamics of spatially flat Einstein-Gauss-Bonnet models in various dimensions: High-dimensional $\Lambda$-term case}

\author{Sergey A. Pavluchenko}
\affiliation{Programa de P\'os-Gradua\c{c}\~ao em F\'isica, Universidade Federal do Maranh\~ao (UFMA), 65085-580, S\~ao Lu\'is, Maranh\~ao, Brazil}

\begin{abstract}
In this paper we perform a systematic study of spatially flat $[(3+D)+1]$-dimensional Einstein-Gauss-Bonnet cosmological models with $\Lambda$-term. We consider models that topologically are 
the product of two flat isotropic subspaces with different scale factors. One of these subspaces is three-dimensional and represents our space and the other is $D$-dimensional and represents 
extra dimensions. We consider no {\it ansatz} of the scale factors, which makes our results quite general. With both Einstein-Hilbert and Gauss-Bonnet contributions in play, $D=3$ and the 
general $D\geqslant 4$ cases have slightly different dynamics due to the different structure of the equations of motion. We analytically study equations of motion in both cases and describe 
all possible regimes with special interest on the realistic regimes. Our analysis suggests that the only realistic regime is the transition from high-energy (Gauss-Bonnet) Kasner regime, which 
is the standard cosmological singularity in that case, to the anisotropic exponential regime with expanding three and contracting extra dimensions.
Availability of this regime allows us to put constraint on the value of Gauss-Bonnet coupling $\alpha$ and the $\Lambda$-term -- this regime appears in two regions on $(\alpha, \Lambda)$ plane:
$\alpha < 0$, $\Lambda > 0$, $\alpha\Lambda \leqslant 1/2$ and $\alpha > 0$, $\alpha\Lambda  \leqslant (3D^2 - 7D + 6)/(4D(D-1))$,
including entire $\Lambda < 0$ region. The obtained bounds are confronted with the restrictions on $\alpha$ and $\Lambda$ from other considerations, like causality, entropy-to-viscosity ratio
in AdS/CFT and others. Joint analysis constraints ($\alpha$, $\Lambda$) even further: $\alpha > 0$, $D \geqslant 2$ with 
$(3D^2 - 7D + 6)/(4D(D-1)) \geqslant \alpha \Lambda  \geqslant - (D+2)(D+3)(D^2 + 5D + 12)/(8(D^2 + 3D + 6)^2)$.
\end{abstract}

\pacs{04.20.Jb, 04.50.-h, 98.80.-k}


\maketitle

\section{Introduction}

It already has been more then a hundred years since the formulation of the General Relativity (GR) by Albert Einstein, but apparently the idea of extra dimensions is even older than that. Indeed, it was Nordstr\"om
who constructed the first ever extra-dimensional model~\cite{Nord1914} in 1914, and this model unified Nordstr\"om's second gravity theory~\cite{Nord_2grav}
with Maxwell's electromagnetism. 
Later in 1915 Einstein introduced General Relativity~\cite{einst}, but still it took almost four years to prove that Nordstr\"om's theory and others were wrong. During the solar eclipse of 1919, the
bending of light near the Sun was measured and the
deflection angle was in perfect agreement with GR, while Nordstr\"om's theory, being scalar gravity, predicted a zeroth deflection angle.

Unlike Nordstr\"om's scalar gravity, his idea about extra dimensions survived, and in 1919 Kaluza proposed~\cite{KK1} a similar model but based on GR: in his model five-dimensional Einstein equations could be decomposed into
four-dimensional Einstein equations
plus Maxwell's electromagnetism. To perform such a decomposition, the extra dimensions should be ``curled'' or compactified into a circle and ``cylindrical conditions'' should be imposed.
Later in 1926,
Klein proposed~\cite{KK2, KK3} a nice quantum mechanical interpretation of this extra dimension and so the theory called Kaluza-Klein was formally formulated. Remarkably, their theory unified all
known interactions at that time. With time, more interactions were known and it became clear that to unify them all, more extra dimensions are needed. Nowadays, one of the promising theories to unify
all interactions is M/string theory.

Presence of the curvature-squared corrections in the Lagrangian of the gravitational counterpart of string theories is one of their distinguishing features.
Scherk and Schwarz~\cite{sch-sch} were the
first to demonstrate the presence of the $R^2$ and
$R_{\mu \nu} R^{\mu \nu}$ terms in the Lagrangian of the Virasoro-Shapiro
model~\cite{VSh1, VSh2}. A presence of curvature-squared term of the $R^{\mu \nu \lambda \rho}
R_{\mu \nu \lambda \rho}$ type was found~\cite{Candelas_etal} in the low-energy limit
of the $E_8 \times E_8$ heterotic superstring theory~\cite{Gross_etal} to match the kinetic term
for the Yang-Mills field. Later it was demonstrated~\cite{zwiebach} that the only
combination of quadratic terms that leads to a ghost-free nontrivial gravitation
interaction is the Gauss-Bonnet (GB) term:

$$
L_{GB} = L_2 = R_{\mu \nu \lambda \rho} R^{\mu \nu \lambda \rho} - 4 R_{\mu \nu} R^{\mu \nu} + R^2.
$$

\noindent This term, first found
by Lanczos~\cite{Lanczos1, Lanczos2} (therefore it is sometimes referred to
as the Lanczos term) is an Euler topological invariant in (3+1)-dimensional
space-time, but not in (4+1) and higher dimensions.
Zumino~\cite{zumino} extended Zwiebach's result on
higher-than-squared curvature terms, supporting the idea that the low-energy limit of the unified
theory might have a Lagrangian density as a sum of contributions of different powers of curvature. In this regard the Einstein-Gauss-Bonnet (EGB) gravity could be seen as a subcase of more general Lovelock
gravity~\cite{Lovelock}, but in the current paper we restrain ourselves with only quadratic corrections and so to the EGB case.

All extra-dimensional theories have one thing in common---it is needed to explain where additional dimensions are ``hiding'', since we do not sense them, at least with the current level of experiments. One of
the ways to hide extra dimensions and to recover four-dimensional physics, is to build a so-called ``spontaneous compactification'' solution. Exact static solutions with the metric chosen as a cross product of a
(3+1)-dimensional manifold and a constant curvature ``inner space'',  were found for the first time in~\cite{add_1}, but the (3+1)-dimensional manifold being Minkowski (the generalization for
a constant curvature Lorentzian manifold was done in~\cite{Deruelle2}).
In the context of cosmology, it is more useful to consider a spontaneous compactification in case with the four-dimensional part given by a Friedmann-Robertson-Walker metric.
In this case it is also natural to consider the size of the extra dimensions being time dependent and not than static. Indeed, in
\cite{add_4} it was exactly demonstrated that in order to have a more realistic model one needs to consider the dynamical evolution of the extra-dimensional scale factor.
In~\cite{Deruelle2}, the equations of motion with time-dependent scale factors were written for arbitrary Lovelock order in the special case of a spatially flat metric (the results were further proven in~\cite{prd09}).
The results of~\cite{Deruelle2} were further analyzed for the special case of 10 space-time dimensions in~\cite{add_10}.
In~\cite{add_8}, the dynamical compactification solutions were studied with the use of Hamiltonian formalism.
More recently, searches for spontaneous  compactifications were made in~\cite{add13}, where
the dynamical compactification of the (5+1) Einstein-Gauss-Bonnet model was considered; in \cite{MO04, MO14} with different metric {\it Ans\"atze} for scale factors
corresponding to (3+1)- and extra-dimensional parts; and in \cite{CGP1, CGP2, CGPT}, where general (e.g., without any {\it Ansatz}) scale factors and curved manifolds were considered. Also, apart from
cosmology, the recent analysis has focused on
properties of black holes in Gauss-Bonnet~\cite{alpha_12, add_rec_1, add_rec_2, addn_1, addn_2} and Lovelock~\cite{add_rec_3, add_rec_4, addn_3, addn_4, addn_4.1} gravities, features of gravitational collapse in these
theories~\cite{addn_5, addn_6, addn_7}, general features of spherical-symmetric solutions~\cite{addn_8}, and many others.

If we want to find exact solutions, the most common {\it Ansatz} used for the scale factor is exponential or power law.
Exact solutions with exponents  for both  the (3+1)- and extra-dimensional scale factors were studied for the first time in~\cite{Is86}, where an exponentially increasing (3+1)-dimensional
scale factor and an exponentially shrinking extra-dimensional scale factor were described.
Power-law solutions have been analyzed  in \cite{Deruelle1, Deruelle2} and more  recently in~\cite{mpla09, prd09, Ivashchuk, prd10, grg10} so that by now there is an  almost complete description of the solutions of this kind
(see also~\cite{PT} for comments regarding physical branches of the power-law solutions).
Solutions with exponential scale factors~\cite{KPT} have also been studied in detail, namely, the models with both variable~\cite{CPT1} and constant~\cite{CST2} volume; the general scheme for
constructing solutions in EGB was developed and generalized for general Lovelock gravity of any order and in any dimensions~\cite{CPT3}. Also, the stability of the solutions was addressed in~\cite{my15} 
(see also~\cite{iv16} for stability of general exponential
solutions in EGB gravity), and it was
demonstrated that only a handful of the solutions could be called ``stable'', while the most of them are either unstable or have neutral/marginal stability, and so additional investigation is
required.

In order to find all possible regimes of Einstein-Gauss-Bonnet cosmology, one needs to go beyond an exponential or power-law {\it Ansatz} and keep the scale factor generic.
We are particularly interested in models that allow dynamical compactification, so that we
consider the metric as the product of a spatially three-dimensional and extra-dimensional parts. In that case the three-dimensional part is ``our Universe'' and we expect for this part to
expand
while the extra-dimensional part should be suppressed in size with respect to the three-dimensional one. In \cite{CGP1} we demonstrated the there existence of phenomenologically
sensible regime when the curvature of the extra dimensions is negative and the Einstein-Gauss-Bonnet theory does not admit a maximally symmetric solution. In this case both the
three-dimensional Hubble parameter and the extra-dimensional scale factor asymptotically tend to the constant values. In \cite{CGP2} we performed a detailed analysis of the cosmological dynamics in this model
with generic couplings. Recently we studied this model in~\cite{CGPT} and demonstrated that, with an additional constraint on couplings, Friedmann-type late-time behavior
could be restored.

The current paper is a spiritual successor of~\cite{my16a, my16b}, where we investigated cosmological dynamics of the vacuum and low-dimensional $\Lambda$-term Einstein-Gauss-Bonnet model. 
In both papers the spatial section is a product of two
spatially flat
manifolds with one of them three-dimensional, which represents our Universe and the other is extra-dimensional. In~\cite{my16a} we considered vacuum model while in~\cite{my16b} and in  
the current paper -- the model with the
cosmological
term. In~\cite{my16a} we demonstrated that the vacuum model has two physically viable regimes -- first of them is the smooth transition from high-energy GB Kasner to low-energy GR Kasner. This regime
appears for $\alpha > 0$ at $D=1,\,2$ (the number of extra dimensions) and for $\alpha < 0$ at $D \geqslant 2$ (so that at $D=2$ it appears for both signs of $\alpha$). The other viable regime is smooth 
transition from high-energy GB
Kasner to anisotropic exponential regime with expanding three-dimensional section (``our Universe'') and contracting extra dimensions; this regime occurs only for $\alpha > 0$ and at $D \geqslant 2$.
In~\cite{my16b} we considered low-dimensional $\Lambda$-term case and it appears that only one of the realistic regimes from the vacuum case is present in low-dimensional $\Lambda$-term case, namely,
the transition from high-energy GB Kasner to anisotropic exponential regime with expanding three-dimensional section (``our Universe'') and contracting extra dimensions; the low-energy GR Kasner is
forbidden in the presence of the $\Lambda$-term so the corresponding transition do not occur. But this is not the only difference -- in $D=1$ $\Lambda$-term case there are no viable regimes at all, making
it pathological (on contrast, vacuum $D=1$ have GB Kasner to GR Kasner viable transition). In $D=2$ we have realistic regime -- GB Kasner to anisotropic exponential solution, but it appear only for $\Lambda > 0$,
so that we do not have viable AdS cosmologies in $D=\{1, 2\}$. So that in this paper we continue the investigation of the $\Lambda$-term case for $D=3$ and general $D \geqslant 4$ cases.
Let us also note that in~\cite{CGP1, CGP2, CGPT} we considered similar model but with both manifolds to be constant (generally non-zero) curvature; the realistic regime in that model has exponential expansion 
of the three-dimensional subspace and constant-size extra dimensions.

Both in~\cite{my16b} and in this paper we consider the cosmological dynamics in the presence of $\Lambda$-term. And in both papers we consider both signs for the value of the cosmological
constant. For a cosmologist, especially physical cosmologist, it could sound blasphemy, but in the high-energy and gravitation theory it is usual pre-requisite. For instance, $\Lambda < 0$ is
needed for a black hole to reach thermal equilibrium with a heat bath~\cite{l1}, or to derive a correct definition of some of the Noether charges~\cite{l2, l3} (see also~\cite{l4}). So that
we conclude that there are enough of reasons to consider both signs of the $\Lambda$-term.

The structure of the manuscript is as follows: first we write down general equations of motion for Einstein-Gauss-Bonnet gravity, then we rewrite them for our {\it Ansatz}. In the following
sections we analyze them for $D=3$ and the general $D\geqslant 4$ cases, considering the $\Lambda$-term case in this paper only. Each case is
followed by a small discussion of the results and properties of this particular case; after considering all cases we discuss their properties, generalities, and differences, compare the limits on $\alpha$
and $\Lambda$ with those from other sources and draw conclusions.

\section{Equations of motion}

As mentioned above, we consider the spatially flat anisotropic cosmological model in Einstein-Gauss-Bonnet gravity with $\Lambda$-term as a matter source (the vacuum model was considered previously in~\cite{my16a}).
The equations of motion for such model include both first and second Lovelock contributions and could easily be derived from the general case (see, e.g.,~\cite{prd09}). We consider flat
anisotropic metric

\begin{equation}\label{metric}
g_{\mu\nu} = \diag\{ -1, a_1^2(t), a_2^2(t),\ldots, a_n^2(t)\};
\end{equation}

\noindent the Lagrangian of this theory has the form

\begin{equation}\label{lagr}
{\cal L} = R + \alpha{\cal L}_2 - 2\Lambda,
\end{equation}

\noindent where $R$ is the Ricci scalar and ${\cal L}_2$,

\begin{equation}
{\cal L}_2 = R_{\mu \nu \alpha \beta} R^{\mu \nu \alpha \beta} - 4,
R_{\mu \nu} R^{\mu \nu} + R^2 \label{lagr1}
\end{equation}

\noindent is the Gauss-Bonnet Lagrangian. Then substituting (\ref{metric}) into the Riemann $R_{\mu \nu \alpha \beta}$ and Ricci $R_{\mu \nu}$ tensors and the scalar in (\ref{lagr}) and (\ref{lagr1}), and varying
(\ref{lagr}) with respect to the metric, we obtain the equations of motion,

\begin{equation}
\begin{array}{l}
2 \[ \sum\limits_{j\ne i} (\dot H_j + H_j^2)
+ \sum\limits_{\substack{\{ k > l\} \\ \ne i}} H_k H_l \] + 8\alpha \[ \sum\limits_{j\ne i} (\dot H_j + H_j^2) \sum\limits_{\substack{\{k>l\} \\ \ne \{i, j\}}} H_k H_l +
3 \sum\limits_{\substack{\{ k > l >  \\   m > n\} \ne i}} H_k H_l
H_m H_n \] - \Lambda = 0
\end{array} \label{dyn_gen}
\end{equation}

\noindent as the $i$th dynamical equation. The first Lovelock term---the Einstein-Hilbert contribution---is in the first set of brackets and the second term---Gauss-Bonnet---is in the second set; $\alpha$
is the coupling constant for the Gauss-Bonnet contribution and we put the corresponding constant for Einstein-Hilbert contribution to unity. Also, since we a consider spatially flat cosmological model, scale
factors do not hold much in the physical sense and the equations are rewritten in terms of the Hubble parameters $H_i = \dot a_i(t)/a_i(t)$. Apart from the dynamical equations, we write down a constraint equation

\begin{equation}
\begin{array}{l}
2 \sum\limits_{i > j} H_i H_j + 24\alpha \sum\limits_{i > j > k > l} H_i H_j H_k H_l = \Lambda.
\end{array} \label{con_gen}
\end{equation}

As mentioned in the Introduction,
we want to investigate the particular case with the scale factors split into two parts -- separately three dimensions (three-dimensional isotropic subspace), which are supposed to represent our world, and the remaining represent the extra dimensions ($D$-dimensional isotropic subspace). So we put $H_1 = H_2 = H_3 = H$ and $H_4 = \ldots = H_{D+3} = h$ ($D$ designs the number of additional dimensions) and the
equations take the following form: the
dynamical equation that corresponds to $H$,

\begin{equation}
\begin{array}{l}
2 \[ 2 \dot H + 3H^2 + D\dot h + \dac{D(D+1)}{2} h^2 + 2DHh\] + 8\alpha \[ 2\dot H \(DHh + \dac{D(D-1)}{2}h^2 \) + \right. \\ \\ \left. + D\dot h \(H^2 + 2(D-1)Hh + \dac{(D-1)(D-2)}{2}h^2 \) +
2DH^3h + \dac{D(5D-3)}{2} H^2h^2 + \right. \\ \\ \left. + D^2(D-1) Hh^3 + \dac{(D+1)D(D-1)(D-2)}{8} h^4 \] - \Lambda=0,
\end{array} \label{H_gen}
\end{equation}

\noindent the dynamical equation that corresponds to $h$,

\begin{equation}
\begin{array}{l}
2 \[ 3 \dot H + 6H^2 + (D-1)\dot h + \dac{D(D-1)}{2} h^2 + 3(D-1)Hh\] + 8\alpha \[ 3\dot H \(H^2 + 2(D-1)Hh + \right .\right. \\ \\ \left. \left. + \dac{(D-1)(D-2)}{2}h^2 \) +  (D-1)\dot h \(3H^2 + 3(D-2)Hh +
\dac{(D-2)(D-3)}{2}h^2 \) + 3H^4 +
\right. \\ \\ \left. + 9(D-1)H^3h + 3(D-1)(2D-3) H^2h^2 +  \dac{3(D-1)^2 (D-2)}{2} Hh^3 + \right. \\ \\ \left. + \dac{D(D-1)(D-2)(D-3)}{8} h^4 \] - \Lambda =0,
\end{array} \label{h_gen}
\end{equation}

\noindent and the constraint equation,

\begin{equation}
\begin{array}{l}
2 \[ 3H^2 + 3DHh + \dac{D(D-1)}{2} h^2 \] + 24\alpha \[ DH^3h + \dac{3D(D-1)}{2}H^2h^2 + \dac{D(D-1)(D-2)}{2}Hh^3 + \right. \\ \\ \left. + \dac{D(D-1)(D-2)(D-3)}{24}h^4\] = \Lambda.
\end{array} \label{con2_gen}
\end{equation}

Looking at (\ref{H_gen}) and (\ref{h_gen}) one can see that for $D\geqslant 4$ the equations of motion contain the same terms, while for $D=\{1, 2, 3\}$ the terms are different (say, for $D=3$ terms with
the $(D-3)$ multiplier are absent and so on) and the dynamics should be different also.
As we mentioned, in this paper we are going to consider only the $\Lambda$-term case; the vacuum case we considered in the previous
paper~\cite{my16a}
while the general case with a perfect fluid with an arbitrary equation of state we as well as the effect of curvature, are going to be considered in the papers to follow. As we also noted, in this
particular paper we consider the $D=3$ and general $D\geqslant 4$ cases -- the low-dimensional $D=\{1, 2\}$ cases were considered in the previous paper~\cite{my16b}.

\section{$D=3$ case}

In this case the equations of motion take the form ($H$-equation, $h$-equation, and constraint correspondingly)

\begin{equation}
\begin{array}{l}
4\dot H + 6H^2 + 6\dot h + 12h^2 + 12Hh + 8\alpha \( 6\dot Hh (H + h) + 3\dot h (H^2 + h^2 + 4Hh) +18H^2h^2 + \right.\\ \left. +18Hh^3 + 3h^4 + 6H^3h\) -\Lambda = 0,
\end{array} \label{D3_H}
\end{equation}

\begin{equation}
\begin{array}{l}
6\dot H + 12H^2 + 4\dot h + 6h^2 + 12Hh + 8\alpha \( 3\dot H (H^2 + 4Hh + h^2) + 6\dot h H (H+h) + 6Hh^3 + \right. \\ \left. + 18H^2h^2 + 18H^3h + 3H^4  \) -\Lambda = 0,
\end{array} \label{D3_h}
\end{equation}

\begin{equation}
\begin{array}{l}
6 H^2 + 18Hh + 6h^2 + 24\alpha (3H^3h + 9H^2h^2 + 3Hh^3 ) = \Lambda.
\end{array} \label{D3_con}
\end{equation}

In the previous paper, dedicated to the low-$D$ case~\cite{my16b}, we solved constraint equation with respect to $h$, because for $D=1$ it is linear with respect to it and for $D=2$ it is quadratic. But
in $D=3$ it is cubic with respect to both $H$ and $h$ while for $D\geqslant 4$ it is fourth order with respect to $H$ and cubic with respect to $h$. So starting from $D=3$ we solve constraint with respect
to $h$ -- by the way, in the paper dedicated to the vacuum case~\cite{my16a}, we did exactly the same. Obviously, both choices -- to solve constraint with respect to $h$ or $H$ -- give the same
results but with different level of complexity.

\begin{figure}
\includegraphics[width=1.0\textwidth, angle=0]{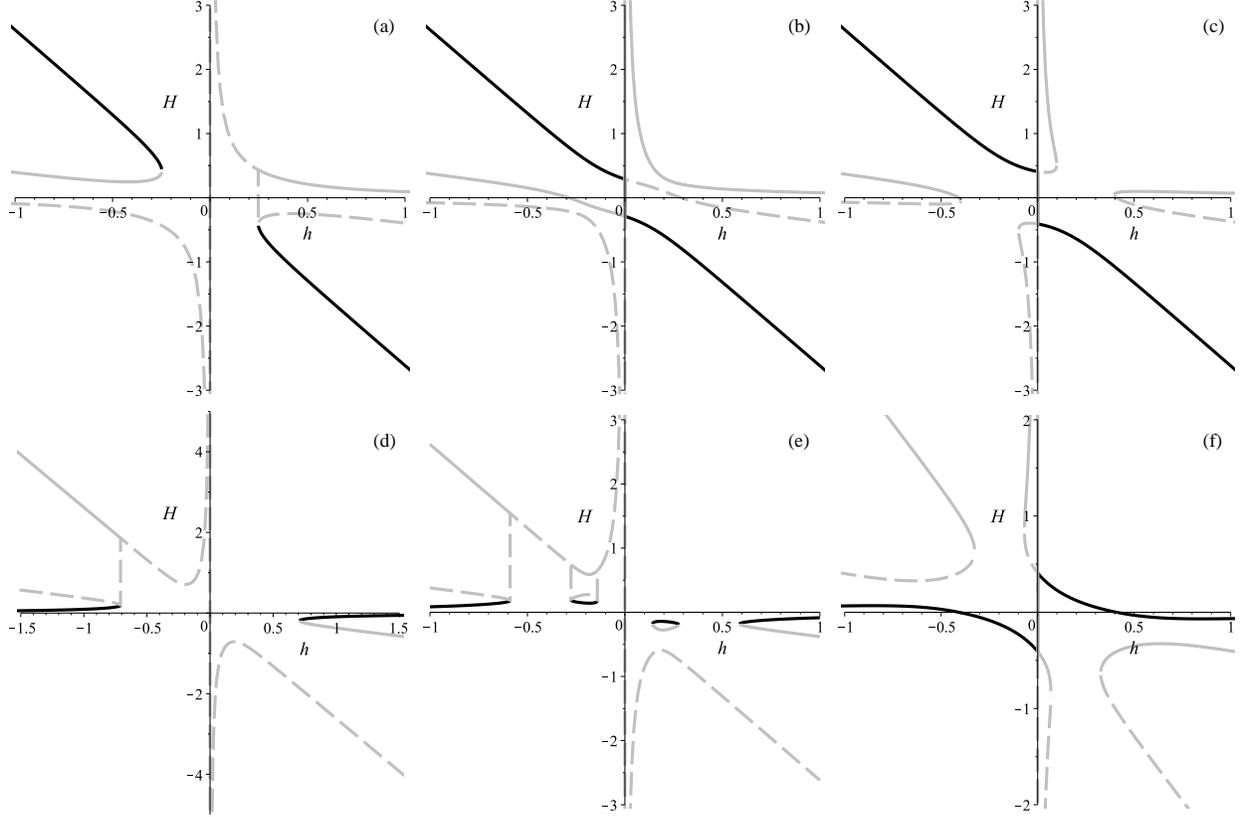}
\caption{$H(h)$ graphs for $D=3$ case: $\alpha < 0$, $\Lambda < 0$ on (a) panel; $\alpha < 0$, $\Lambda > 0$ with $\alpha \Lambda \geqslant -5/8$ on (b) and with $\alpha \Lambda < -5/8$ on (c) panels;
$\alpha > 0$, $\Lambda < 0$ with $\alpha \Lambda > -1/8$ on (d) and with $\alpha \Lambda < -1/8$ on (e) panels; $\alpha > 0$, $\Lambda > 0$ on (f) panel. Different branches presented by different linestyle/color
combination: $H_1$ by solid black line, $H_2$ by solid grey and $H_3$ by dashed grey
(see the text for more details).}\label{D3_1}
\end{figure}

Solving (\ref{D3_con}) with respect to $H$ gives us three roots, $H_1$, $H_2$, and $H_3$ whose expressions are too long to write them down here.
Consideration of the discriminant of (\ref{D3_con}) (with respect to $H$) gives us two values for $\alpha\Lambda$ where the behavior qualitatively changes: $\alpha\Lambda = \{-5/8, -1/8\}$.
We presented $H(h)$ graphs in Figure~\ref{D3_1}. There on (a)--(c) panels we presented the results for $\alpha < 0$ and on (d)--(f) -- for $\alpha > 0$ cases. In particular, (a) panel represents $\alpha < 0$, $\Lambda < 0$
case, (b) panel $\alpha < 0$, $\Lambda > 0$ with $\alpha \Lambda \geqslant -5/8$, and (c) panel $\alpha < 0$, $\Lambda > 0$ with $\alpha \Lambda < -5/8$. On $\alpha > 0$ domain, (d) panel corresponds to $\Lambda < 0$ with
$\alpha \Lambda > -1/8$ while (e) panel -- to $\alpha \Lambda < -1/8$ cases. Finally, on (f) panel we presented $H(h)$ for $\alpha > 0$, $\Lambda > 0$ case. Let us comment these cases a bit.

From Fig.~\ref{D3_1}(a) we can see that for $\alpha < 0$, $\Lambda < 0$ neither $H_1$ nor $H_2$ have $h\to 0$ asymptotes -- only $H_3$ has. So that both $H_1$ and $H_2$ have limited domain of definition while $H_3$ defined everywhere except $h=0$, and
have a removable discontinuity at some $h > 0$. At this point the solution do not follow the branch, but ``jumps'' to another branch -- from $H_3$ to $H_2$. Similarly the solution smoothly jumps from $H_1$ to $H_2$ at
$h<0$, $H>0$ and from $H_3$ to $H_1$ at $h>0$, $H<0$.
The case $\alpha < 0$, $\Lambda > 0$ with $\alpha \Lambda > -5/8$, presented in Fig.~\ref{D3_1}(b), has all branches with finite or infinite limits at $h\to 0$, but
all of them are directional: $\lim\limits_{h\to 0+0} H \ne \lim\limits_{h\to 0-0} H$. Also in this case all three branches are defined everywhere. Similarly to the previous case, we have solutions which change branches while
evolving -- $H_1$ branch shifts to $H_3$ at $h=0$ and $H_2$ shifts to $H_1$ at the same point -- we describe it more while discussing the regimes and their abundances.
For $\alpha < 0$, $\Lambda > 0$ with $\alpha \Lambda < -5/8$ case presented in Fig.~\ref{D3_1}(c) we again have finite or infinite limits at $h\to 0$, they also are directional and in this case the domain of definition is not
entire $h \in \mathds{R}$ for all branches; also we have more abundant changes between the branches.

In Figs.~\ref{D3_1}(d)--(f) we presented $\alpha > 0$ cases. The (d) panel corresponds to $\Lambda < 0$ with $\alpha \Lambda > -1/8$ while (e) -- the same but with $\alpha \Lambda < -1/8$. 
One can see that the difference
between these two is the existence of additional isolated regions of definition for $H_1$ and $H_2$ branches. They give some additional regimes (and ``redistributing'' evolution between $H_2$ and $H_3$ for $h<0$) but since
they are isolated, they cannot give rise to viable regimes -- we will check that when describing regimes. Finally, in Fig.~\ref{D3_1}(f) we presented $H(h)$ curves for $\alpha > 0$, $\Lambda > 0$ case. All these three cases also
have their share of branch changing and while describing the actual regimes we pay close attention to these changes.

Looking at Fig.~\ref{D3_1} we can make predictions for the existence of isotropic solutions -- indeed, as isotropy requires $H=h$, we can tell from Fig.~\ref{D3_1} that we have one isotropic solution for (a) and (f)
panels, two for (b) panel and no solutions for the rest of the panels.

Our analysis of the previous cases~\cite{my16a, my16b} revealed that we expect to encounter exponential solutions -- isotropic and anisotropic. Below we find the condition for both kinds of exponential solutions to exist.
Isotropic solutions are governed by the equation

\begin{equation}
\begin{array}{l}
360\alpha H^4 + 30 H^2 - \Lambda = 0,
\end{array} \label{D.3_iso_1}
\end{equation}

\noindent and one can see that it is a biquadratic equation with respect to $H$. So that for solutions to exist we need not only positivity of the discriminant, but positivity of the roots. Then, skipping the derivation,
we can see that for $\alpha > 0$, $\Lambda < 0$ there are no isotropic solutions, for $\alpha > 0$, $\Lambda > 0$ as well as for $\alpha < 0$, $\Lambda < 0$ there is one, and for $\alpha < 0$, $\Lambda > 0$,
$\alpha\Lambda < -5/8$ there are no solutions while for $\alpha < 0$, $\Lambda > 0$, $\alpha\Lambda > -5/8$ there are two.

The existence of anisotropic exponential solutions is governed by the following equation

\begin{equation}
\begin{array}{l}
20736 \xi^4 - 17280 \xi^3 + (4032\zeta + 2736)\xi^2 - (528\zeta + 216)\xi + (4\zeta^2 + 12\zeta + 9) = 0,
\end{array} \label{D.3_aniso_1}
\end{equation}

\noindent where $\xi = \alpha h^2$ and $\zeta = \alpha\Lambda$.
Its discriminant $\Delta = 328683126924509184 (8\zeta + 5) (8\zeta - 3) (2\zeta - 1)^4$ clearly gives us values for $\zeta$ which separate regions with different root numbers. So that for $\zeta < \zeta_1 = -5/8$ we have
four roots -- two for $\xi < 0$ (and so $\alpha < 0$ since $\xi = \alpha h^2$) and two for $\xi > 0$. At $\zeta = \zeta_1$ negative roots coincide and for $0 > \zeta > \zeta_1$ they disappear while positive roots remain.
For positive $0 < \zeta < \zeta_2 = 3/8$ we have two positive roots, for $\zeta = \zeta_2$ its number increase to three and for $\zeta_2 < \zeta< \zeta_3 = 1/2$ its further increase to four. At $\zeta = \zeta_3$ it reduces
to two and finally for $\zeta > \zeta_3$ they disappear -- so that for $\zeta > \zeta_3 = 1/2$ there are no anisotropic exponential solutions -- the same situation we have in $D=2$ $\Lambda$-term case~\cite{my16b}.

Before describing the resulting $\dot h(h)$ and $\dot H(h)$ graphs and the regimes, let us make a note on definitions. They are similar to those we used in previous papers~\cite{my16a, my16b}: 
we denote power-law regimes
with $K_i$ and the index $i$ corresponds to the sum of Kasner exponents $\sum p$: $\sum p=1$ for GR ($K_1$) and $\sum p=3$ for GB ($K_3$). In this study we do not have $K_1$ for reasons discussed in~\cite{my16b};
formally we should not have
$K_3$ either but in the GB regime $H_i \ggg \Lambda$ so we can treat $\Lambda$-term regime as asymptotically flat
(see~\cite{my16b} for details). Apart from the power-law solutions we could have exponential solutions -- we denote them as $E$ with indices indicating their features -- say,
$E_{iso}$ is isotropic exponential solution; if there are two different isotropic exponential solutions, we denote them as $E_{iso}^1$ and $E_{iso}^2$. Anisotropic exponential solutions are denoted just as $E$ and the
indices enumerate the solutions (e.g., $E_1$, $E_2$ etc).

\begin{figure}
\includegraphics[width=1.0\textwidth, angle=0]{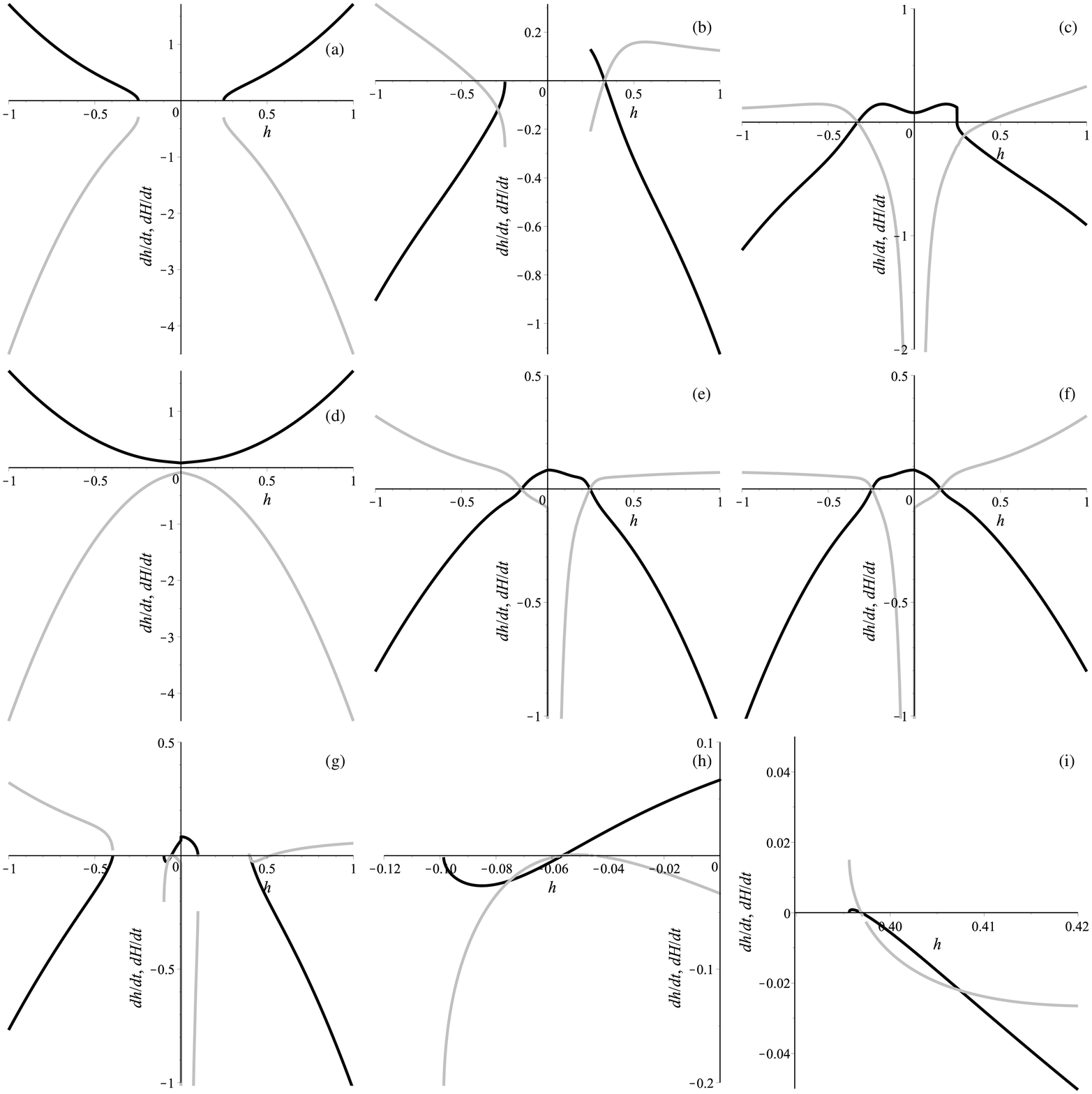}
\caption{$\dot h(h)$ (black) and $\dot H(h)$ (grey) curves for $\alpha < 0$ $D=3$ $\Lambda$-term cases: $\Lambda < 0$ on (a)--(c) panels -- $H_1$ branch on (a), $H_2$ branch on (b) and $H_3$ branch on (c); $\Lambda > 0$
with $\alpha \Lambda \geqslant -5/8$ on (d)--(f) panels -- $H_1$ branch on (d), $H_2$ branch on (e) and $H_3$ branch on (f); $\Lambda > 0$ with $\alpha \Lambda < -5/8$ on (g)--(i) panels -- $H_2$ on all three panels --
large scale
on (g) panel and detailed features for $h<0$ on (h) and for $h>0$ on (i) panels
(see the text for more details).}\label{D3_2}
\end{figure}

\begin{figure}
\includegraphics[width=1.0\textwidth, angle=0]{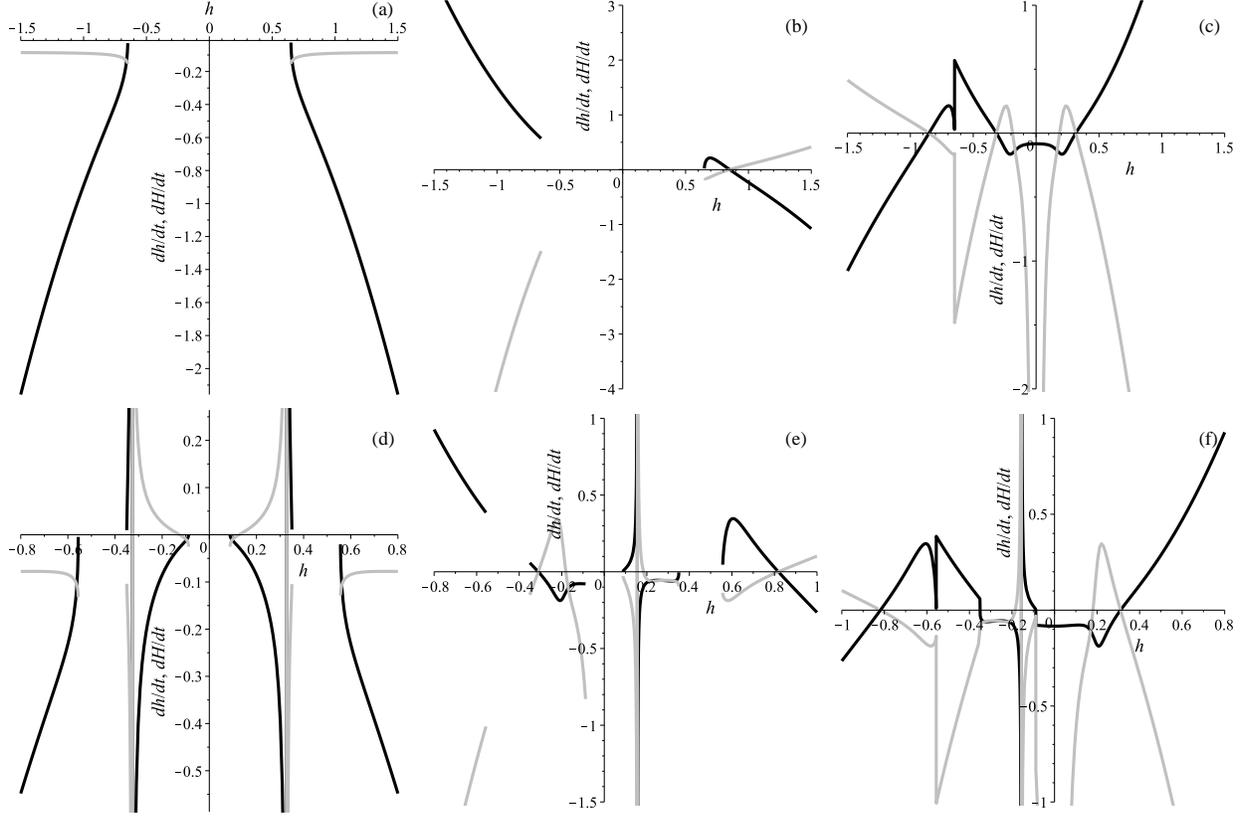}
\caption{$\dot h(h)$ (black) and $\dot H(h)$ (grey) curves for $\alpha > 0$, $\Lambda < 0$ $D=3$ $\Lambda$-term cases: $\zeta \leqslant \zeta_{cr}$ on (a)--(c) panels and $\zeta > \zeta_{cr}$ on (d)--(f) panels.
On (a) and (d) panels we presented $H_1$ branch, on (b) and (e) -- $H_2$ and on (c) and (f) -- $H_3$
(see the text for more details).}\label{D3_3}
\end{figure}

\begin{figure}
\includegraphics[width=1.0\textwidth, angle=0]{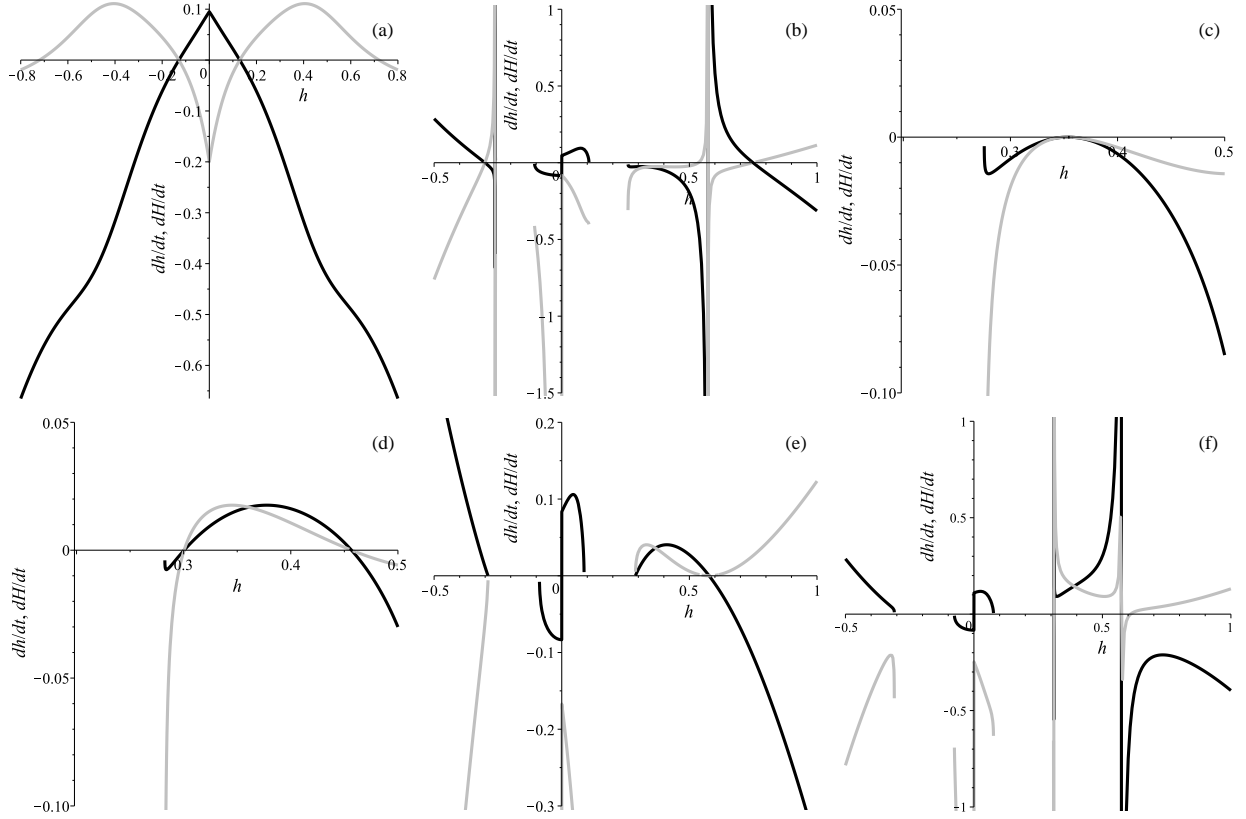}
\caption{$\dot h(h)$ (black) and $\dot H(h)$ (grey) curves for $\alpha > 0$, $\Lambda > 0$ $D=3$ $\Lambda$-term cases: $H_1$ on (a) panel and $H_2$ on the remaining: $\alpha\Lambda < 3/8$ on (b), detail of
$\alpha\Lambda = 3/8$ on (c), the same detail for $1/2 > \alpha\Lambda > 3/8$ on (d), $\alpha\Lambda = 1/2$ on (e) and $\alpha\Lambda > 1/2$ on (f)
(see the text for more details).}\label{D3_4}
\end{figure}

For the further analysis we solve (\ref{D3_H})--(\ref{D3_h}) with respect to $\dot h$ and $\dot H$ and substitute $H_i(h)$ branches into them; as a result we have three $\dot h_i(h)$ and $\dot H_i(h)$ curves which
correspond to three branches. The expressions for $\dot h_i(h)$ and $\dot H_i(h)$ are too lengthy so we do not write them down but present them in Figs.~\ref{D3_2}--\ref{D3_4} for different cases.

In Fig.~\ref{D3_2} we presented $\dot h(h)$ (in black) and $\dot H(h)$ (in gray) curves for $\alpha < 0$ $D=3$ $\Lambda$-term cases: $\Lambda < 0$ on (a)--(c) panels -- $H_1$ branch on (a), $H_2$ branch on (b) and $H_3$ branch
on (c); $\Lambda > 0$ with $\alpha \Lambda \geqslant -5/8$ on (d)--(f) panels -- $H_1$ branch on (d), $H_2$ branch on (e) and $H_3$ branch on (f); $\Lambda > 0$ with $\alpha \Lambda < -5/8$ on (g)--(i) panels -- $H_2$ on all three
panels -- large scale on (g) panel and detailed features for $h<0$ on (h) and for $h>0$ on (i) panels. Let us have a closer look on them and with additional use of Fig.~\ref{D3_1} describe all possible regimes.

Fist let us describe $\alpha < 0$, $\Lambda < 0$ regimes -- $H_1$ branch (see Fig.~\ref{D3_2}(a)) for $h < 0$ has $K_3$ as past asymptote and switch to $H_2$ in future with another $K_3$ asymptote (see Fig.~\ref{D3_1}(a)).
Similar situation is with $h>0$ -- there $H_1$ branch describe future $K_3$ asymptote and the past is $K_3$ from $H_3$ branch.  Another branch, $H_2$, presented in Fig.~\ref{D3_2}(b), and for $h<0$ it is complimentary
for $H_1$ branch -- they form $K_3 \to K_3$ transition.
For $h>0$ this branch has isotropic exponential solution as a future attractor and $K_3$ as a past attractor; additional past attractor is another $K_3$ from $H_3$. Finally $H_3$ branch, presented in Fig.~\ref{D3_2}(c),
is smooth in ($h<0$, $H<0$) and
has two parts in $h>0$ -- one part with $H>0$ and another with $H<0$ (see Fig.~\ref{D3_1}(a)). So for $h<0$ we have two regimes $E_{iso} \to K_3$ but
with two different $K_3$ -- one at $h\to 0$ and another at $h\to -\infty$. For $h>0$, the part with $H<0$ serves as a past $K_3$ asymptote for $K_3\to K_3$ transition from $H_3$ to $H_1$ while 
$H>0$ part is one of the past $K_3$
asymptotes for isotropic exponential solution from $H_2$. To summarize, in $\alpha < 0$, $\Lambda < 0$ case there are no viable regimes -- we have only $K_3\to K_3$ and $K_3 \leftrightarrow E_{iso}$.

The second case, $\alpha < 0$, $\Lambda > 0$, $\alpha \Lambda \geqslant -5/8$, is presented in Figs.~\ref{D3_2}(d)--(f) and Fig.~\ref{D3_1}(b). First branch, $H_1$, presented in Fig.~\ref{D3_2}(d). One can see that both
parts -- $h<0$ and $h>0$ -- have $K_3$ at $h\to\pm\infty$ and shift to other branches at $h=0$. One also cannot miss that $H_2$ (Fig.~\ref{D3_2}(e)) and $H_3$ (Fig.~\ref{D3_2}(f)) branches have ``mirror-symmetry'' with respect
to $h=0$. One can see that
there are two exponential solutions, both of them are isotropic, we denote the solution at smaller $|h|$ as $E_{iso}^1$ and at larger -- $E_{iso}^2$. So that $H_2$ in $h<0$ domain has $E_{iso}^1$ as a past asymptote and $K_3$
as a future (another future $K_3$ asymptote is on $H_1$ branch). In $h>0$ domain $H_2$ branch has $E_{iso}^2$ as future asymptote and two different $K_3$ -- for $h\to 0$ and $h\to +\infty$. The description for $H_3$ is
similar to the description of $H_2$ with the mentioned above ``mirror symmetry'' kept in mind -- $E_{iso}^1$ is replaced with $E_{iso}^2$ and vice versa, past asymptote with future and vice versa and so on. To summarize,
in this  $\alpha < 0$, $\Lambda > 0$, $\alpha \Lambda \geqslant -5/8$ case we also do not have viable regimes -- we have two different isotropic exponential solutions and different $K_3$ as both past and future asymptotes.
Let us note that as $\alpha \Lambda$ decreases, the separation between $E_{iso}^1$ and $E_{iso}^2$ reduces as well, so for $\alpha \Lambda = -5/8$ two isotropic exponential solutions coincide.

With further decrease of $\alpha \Lambda$ -- for $\alpha \Lambda < -5/8$ -- the situation changes and the corresponding $H(h)$ curves are presented in Fig.~\ref{D3_1}(c).
$H_1$ branch looks exactly the same as in the previous case -- Fig.~\ref{D3_2}(d), so do the regimes -- $K_3$ as a past asymptote for $h<0$ and future asymptote for $h>0$ and transition to another branch at $h=0$.
Another branch, $H_2$, is presented in Figs.~\ref{D3_2}(g)--(i) -- large-scale on (g) panel and detailed two exponential solution -- for $h<0$ on (h) and for $h>0$ on (i). So for $h<0$ on the ``outer'' part we have $K_3$ as a
past asymptote after branch changing, $h<0$ ``inner'' part has anisotropic exponential solution $E_1$ as a past asymptote and branch changing at $h=0$ and at some $h_1< 0$. Regimes for $h>0$ include $K_3$ as a past
asymptote and regime changing at some $h_1>0$, and anisotropic exponential solution $E_2$ as a future asymptote with $K_3$ and branch switching at some $h_0 > 0$.
The final branch, $H_3$, has the same ``mirror symmetry'' as described above so all the regimes could be obtained from the description of $H_2$ with appropriate replacements, similar to the procedure for $H_2$
($h\to -h$, $H\to -H$, past $\to$ future etc). Combined analysis shows that in this case there are no isotropic exponential solutions but there are anisotropic. But all of them have 
$H>0$, $h>0$ (so both three- and extra-dimensional spaces are expanding), and for that reason we cannot call them viable.

Our analysis continues with $\alpha > 0$ cases and we start with $\Lambda < 0$ presented in Fig.~\ref{D3_3}. There on the upper row we presented $\zeta \leqslant \zeta_{cr}$ ($\alpha\Lambda \leqslant -1/8$)
while on the bottom row -- $\zeta > \zeta_{cr}$. On (a) and (d) panels we presented $H_1$ branch, on (b) and (e) -- $H_2$ and on (c) and (f) -- $H_3$. Comparing panels (e) and (f) from Fig.~\ref{D3_1}, we can see the
difference between these two cases -- for $\zeta > \zeta_{cr}$ we have two additional singular regimes. So for $H_1$ we have $K_3$ as a future (for $h<0$) or past (for $h>0$) asymptote of another regime and for
$\zeta > \zeta_{cr}$ we additionally have $nS\to nS$ transition with part of this transition happening on $H_3$ (for $h<0$) and $H_2$ (for $h>0$) branches. For $H_2$ we have $K_3$ as a past asymptote for $h<0$ and
anisotropic exponential solution $E_2$ as a future asymptote for $h>0$. In addition for $\zeta > \zeta_{cr}$ we have part of $nS\to nS$ transition for $h>0$ and another anisotropic exponential solution $E_1$
(future asymptote) is shifted from $H_3$ to $H_2$ for $h<0$. Finally for $H_3$ we have anisotropic exponential solution $E_2$ as past asymptote and anisotropic exponential solution $E_1$ as future asymptote for $h<0$
and anisotropic exponential solution $E_1$ as past asymptote for $h>0$. For $\zeta > \zeta_{cr}$ anisotropic exponential solution $E_1$ is shifted to $H_2$ and additional part of $nS\to nS$ transition is introduced.
In this $\alpha > 0$, $\Lambda < 0$ case we have viable $K_3 \to E_{1, 2}$ regimes (transition from Gauss-Bonnet Kasner regime to anisotropic exponential expansion) and  they occur regardless of $\alpha\Lambda \leqslant -1/8$
or $\alpha\Lambda > -1/8$.

And the final case to consider is $\alpha > 0$, $\Lambda > 0$ which is presented in Fig.~\ref{D3_4}. There on (a) panel we presented the behavior of $H_1$ and it remains the same for all values of $\alpha\Lambda$ for this
case. We can see that there are isotropic exponential solution as past asymptote and $K_3$ as future asymptote for $h<0$ and opposite behavior -- future asymptote for isotropic solution and past for $K_3$ -- for $h>0$. Both halves
change the branch at $h=0$.
On the remaining panels of Fig.~\ref{D3_4} we presented the behavior for $H_2$. On (b) panel we presented the behavior for $\alpha\Lambda < 3/8$: on the outmost part of $h<0$ one can see anisotropic exponential solution
$E_1$ as future asymptote and $K_3$ and $nS$ as past asymptotes on $h\to -\infty$ and the boundary of the region respectively. On the innermost part of $h<0$ we can see $K_3$ as past asymptote and branch change on the
boundary. The innermost part of $h>0$ have branch changes on both $h=0$ and on the boundary. This way for $H_2$ we have $K_3$ for $h\to 0-0$ but no regime for $h\to 0+0$ (since it is branch change), so $K_3$ is reached
directionally. Finally, the outmost part of $h>0$ demonstrates $nS\to nS$ transition and then $nS\to E_2$ and $K_3 \to E_2$, with growth of $h$. With increase of $\alpha\Lambda$ the area between two nonstandard
singularities begins to change -- both $\dot h(h)$ and $\dot H(h)$ increase and at $\alpha\Lambda = 3/8$ they ``touch'' zero, as depicted in Fig.~\ref{D3_4}(c). This creates anisotropic exponential solution $E_{3, 4}$ which
is a past asymptote for one of nonstandard singularities and future asymptote -- for another. With further increase of $1/2 > \alpha\Lambda > 3/8$ the $\dot h(h)$ and $\dot H(h)$ curves further increase and form two
anisotropic exponential solutions $E_3$ and $E_4$, adding exotic $E_3 \to E_4$ transition to the described above picture (see Fig.~\ref{D3_4}(d)). Finally at $\alpha\Lambda = 1/2$, which is presented in Fig.~\ref{D3_4}(e),
drastic changes occur -- anisotropic exponential solution $E_1$ from $h<0$ and nonstandard singularities from $h>0$ disappear, leaving us with $E_3 \to E_4$ and $K_3 \to E_4$ regimes. At $\alpha\Lambda > 1/2$ both singularities
are back while anisotropic exponential solutions  disappear, leaving us with $nS\to nS$ and $K_3 \to nS$. So that in  $\alpha > 0$, $\Lambda > 0$ case we have two viable regimes for $\alpha\Lambda < 1/2$ ($K_3 \to E_1$
at $h<0$ and $K_3 \to E_2$ at $h>0$\footnote{These two regimes are a bit different -- $E_1$ has $h<0$ and $H>0$ while $E_2$ has $h>0$ and $H<0$ but since both spaces are three-dimensional, we can just claim expanding
one as ``our Universe'' and contracting one as extra dimensions, so we do not discriminate between them. Of course in $D\ne 3$ cases this situation cannot appear.}), one viable regime for $\alpha\Lambda = 1/2$
($K_3 \to E_2$) and no viable regimes for $\alpha\Lambda > 1/2$. Let us note that this is quite similar to what we saw in $D=2$ case~\cite{my16b}.

\begin{table}
\begin{center}
\caption{Summary of $D=3$ power-law $\Lambda$-term regimes.}
\label{D.3.0}
  \begin{tabular}{|c|c|c|c|c|c|c|c|c|c|c|}
    \hline
\multirow{2}{*}{Branch} & \multirow{2}{*}{$\alpha$} & \multicolumn{3}{c|}{$h\to +\infty$} & \multicolumn{3}{c|}{$h\to -\infty$} & \multicolumn{3}{c|}{$h\to 0$}  \\ \cline{3-11}
& & $p_H$ & $p_h$ & $\sum p$ & $p_H$ & $p_h$ & $\sum p$ & $p_H$ & $p_h$ & $\sum p$ \\
    \hline
\multirow{2}{*}{$H_1$} & $\alpha > 0$ & 0 & 1 & 3 & 0 & 1 & 3 & 1 & 0 & 3 \\ \cline{2-11}
&     $\alpha < 0$ & $p_1$ & $p_2$ & 3 & $p_1$ & $p_2$ & 3 & 1 & 0 & 3 \\ \cline{1-11}
\multirow{2}{*}{$H_2$} & $\alpha > 0$ & $p_2$ & $p_1$ & 3 & $p_1$ & $p_2$ & 3 & 1 & 0 & 3 \\ \cline{2-11}
& $\alpha < 0$ & 0 & 1 & 3 & $p_2$ & $p_1$ & 3 & 1 & 0 & 3 \\ \cline{1-11}
\multirow{2}{*}{$H_3$} & $\alpha > 0$ & $p_1$ & $p_2$ & 3 & $p_2$ & $p_1$ & 3 & 1 & 0 & 3 \\ \cline{2-11}
& $\alpha < 0$ & $p_2$ & $p_1$ & 3 & 0 & 1 & 3 & 1 & 0 & 3 \\
      \hline
  \end{tabular}
\end{center}
\end{table}

\begin{table}
\begin{center}
\caption{Summary of nontrivial $D=3$ $\alpha < 0$ $\Lambda$-term regimes.}
\label{D.3a}
  \begin{tabular}{|c|c|c|c|c|}
    \hline
   $\alpha, \Lambda$ &  Branch & \multicolumn{2}{c|}{Conditions} & Regimes  \\
    \hline
\multirow{6}{*}{$\Lambda < 0$} & $H_1$ & $h<0$ & $h < h_0$ & $K_3 \to K_3 (H_2, h<0)$ \\ \cline{2-5}
&\multirow{3}{*}{$H_2$} & $h<0$ & $h < h_0$ & $K_3 (H_1, h<0) \to K_3$ \\ \cline{3-5}
& & \multirow{2}{*}{$h>0$} & $h_{e} > h > h_0$ & $K_3 (H_3, h>0) \to E_{iso}$ \\ \cline{4-5}
& & & $h > h_e$ & $K_3 \to E_{iso}$ \\ \cline{2-5}
&\multirow{2}{*}{$H_3$} & \multirow{2}{*}{$h>0$} & $h < h_1$ & $K_3 \to E_{iso} $ \\ \cline{4-5}
& & & $h > h_1$ & $K_3 \to K_3 (H_1, h>0) $ \\ \cline{1-5}
\multirow{6}{*}{$\Lambda>0, \alpha\Lambda \geqslant -5/8$} & $H_1$ & \multicolumn{2}{c|}{$h<0$} & $K_3 \to E_{iso}^1 (H_3, h>0)$ \\ \cline{2-5}
& \multirow{3}{*}{$H_2$} & \multicolumn{2}{c|}{$h<0$} & $E_{iso}^1 \to K_3$ \\ \cline{3-5}
& & \multirow{2}{*}{$h>0$} & $h<h_{e, 2}$ & $K_3 \to E_{iso}^2$ \\ \cline{4-5}
& & & $h>h_{e, 2}$ & $K_3 \to E_{iso}^2$ \\ \cline{2-5}
& \multirow{2}{*}{$H_3$} & \multirow{2}{*}{$h>0$} & $h<h_{e, 1}$ & $K_3 (H_1, h<0) \to E_{iso}^1$ \\ \cline{4-5}
& & & $h>h_{e, 1}$ & $K_3 \to E_{iso}^1$ \\ \cline{1-5}
\multirow{10}{*}{$\Lambda>0, \alpha\Lambda < -5/8$} & $H_1$ & \multicolumn{2}{c|}{$h<0$} & $K_3 \to E_1 (H_3, h>0)$ \\ \cline{2-5}
& \multirow{5}{*}{$H_2$} & \multicolumn{2}{c|}{$h<0$} & $E_2 (H_3, h<0) \to K_3$ \\ \cline{3-5}
& & \multirow{3}{*}{$h>0$} & $h<h_1$ & $K_3 \to E_1 (H_3, h>0)$ \\ \cline{4-5}
& & & $h_{e, 2}>h>h_0$ & $K_3 \to E_2$ \\ \cline{4-5}
& & & $h>h_{e, 2}$ & $K_3 \to E_2$ \\ \cline{2-5}
& \multirow{3}{*}{$H_3$} & \multirow{4}{*}{$h>0$} & $h<h_{e, 1}$ & $K_3 (H_1, h<0) \to E_1$ \\ \cline{4-5}
& & & $h_1>h>h_{e, 1}$ & $K_3 (H_2, h>0) \to E_1$ \\ \cline{4-5}
& & & $h>h_0$ & $K_3 \to E_2 (H_2, h>0)$ \\
      \hline
  \end{tabular}
\end{center}
\end{table}

The $H_3$ branch is symmetric (in the described above sense) to $H_2$ -- past and future asymptotes are interchanged as well as signs for $H$ and $h$ and so on, so we are not giving separate description of $H_3$ regimes.
Also due to this symmetry, all anisotropic exponential solutions of $H_3$ are past asymptotes so it do not give rise to any viable regimes.

It is also useful to provide an analysis in $\{p_H, p_h\}$ coordinates -- in Kasner exponents space. They are defined as follows -- $p_H = - \dot H/H^2$ and $p_h = - \dot h/h^2$, so we substitute the expressions
for $\dot H(h)$ and $\dot h(h)$ as well as $H(h)$ for individual branch and obtain expression for Kasner exponents. Again, they are very large so we do not write them down but perform and analysis of all regimes
in $\{p_H, p_h\}$ coordinates -- the same with what we have done in~\cite{my16a, my16b}. The analysis proved that we properly described all the regimes and that at $h\to\pm\infty$ (and some of $h\to 0$ -- which we claimed
during the description) are really
Gauss-Bonnet Kasner regimes. We provide the limiting values for $p_h$ and $p_H$ in Table~\ref{D.3.0} . In there we used the following notations for two particular values of Kasner exponents:
$p_1 = (34\sqrt{5} - 76)/(123 - 55\sqrt{5}) \approx 1.618$ and $p_2 = -(34\sqrt{5} + 76)/(123 + 55\sqrt{5}) \approx -0.618$.

We already mentioned the ``mirror symmetry'' between $H_2$ and $H_3$ branches--- we often detected it while describing $\dot h(h)$ and $\dot H(h)$ and the regimes. One can see from Fig.~\ref{D3_1} that $H(h)$ curves also
have certain symmetry -- with just $h>0$ or $H>0$ half the entire picture could be recovered. For this reason, when collecting regimes for $\alpha < 0$ cases, we limit ourselves with just $H>0$ regimes. This way we
discard contracting
isotropic regimes and only expanding remain; discarded regimes are ``time-reversed'' of the remaining ones. For $\alpha > 0$ we keep only regimes which involve stable (future asymptote) exponential solutions, as
their unstable counterparts (past asymptotes) could be retrieved just by ``time-reversal''.

\begin{table}
\begin{center}
\caption{Summary of nontrivial $D=3$ $\alpha > 0$ $\Lambda$-term regimes.}
\label{D.3b}
  \begin{tabular}{|c|c|c|c|c|}
    \hline
   $\alpha, \Lambda$ &  Branch & \multicolumn{2}{c|}{Conditions} & Regimes  \\
    \hline
\multirow{6}{*}{$\Lambda < 0$} & $H_1$ & $h>0$ & $h>h_0$ & $K_3 \to E_2 (H_2, h>0)$ \\ \cline{2-5}
& \multirow{3}{*}{$H_2$} & $h<0$ & $h>h_0$ & $K_3 \to E_1 (H_3, h<0)$ \\ \cline{3-5}
& & \multirow{2}{*}{$h>0$} & $h_{e, 2} > h > h_0$ & $K_3 (H_1, h>0) \to E_2$ \\ \cline{4-5}
& & & $h>h_{e, 2}$ & $K_3 \to E_2$ \\ \cline{2-5}
& \multirow{2}{*}{$H_3$} & \multirow{2}{*}{$h<0$} & $h_{e, 1} > h > h_0$ & $K_3 (H_2,h<0) \to E_1$ \\ \cline{4-5}
& & & $h>h_{e, 1}$ & $K_3 \to E_1$ \\ \cline{1-5}
\multirow{2}{*}{$\Lambda > 0$} & \multirow{2}{*}{$H_1$} & \multirow{2}{*}{$h>0$} & $h<h_e$ & $K_3 (H_2, h<0) \to E_{iso}$ \\ \cline{4-5}
& & & $h>h_e$ & $K_3 \to E_{iso}$\\ \cline{1-5}
\multirow{6}{*}{$\Lambda > 0$, $\alpha\Lambda < 3/8$} & \multirow{15}{*}{$H_2$} & \multirow{3}{*}{$h < 0$} & $h<h_{e, 1}$ & $K_3 \to E_1$ \\ \cline{4-5}
& & & $h_1>h>h_{e, 1}$ & $nS\to E_1$ \\ \cline{4-5}
& & & $h>h_0$ & $K_3 \to E_{iso} (H_1, h>0)$ \\ \cline{3-5}
& & \multirow{12}{*}{$h > 0$} & $h_2 > h > h_1$ & $nS \to nS$ \\ \cline{4-5}
& & & $h_{e, 2} > h > h_2$ & $nS \to E_2$ \\ \cline{4-5}
& & & $h>h_2$ & $K_3 \to E_2$\\ \cline{1-1} \cline{4-5}
\multirow{2}{*}{$\Lambda > 0$, $\alpha\Lambda = 3/8$} &  &  & $h_{e 3,4} > h > h_1$ & $E_{3, 4} \to nS$ \\ \cline{4-5}
& & & $h_2 > h > h_{3, 4}$ & $nS \to E_{3, 4}$ \\ \cline{1-1} \cline{4-5}
\multirow{3}{*}{$\Lambda > 0$, $1/2 > \alpha\Lambda > 3/8$} &  &  & $h_{e, 3} > h > h_1$ & $E_3 \to nS$ \\ \cline{4-5}
& & & $h_{e, 4} > h > h_{e, 3}$ & $E_3 \to E_4$ \\ \cline{4-5}
& & & $h_2 > h > h_{e, 4}$ & $nS \to E_4$ \\ \cline{1-1} \cline{4-5}
\multirow{2}{*}{$\Lambda > 0$, $\alpha\Lambda = 1/2$} &  &  & $h_{e, 2} > h > h_1$ & $nS \to E_2$\\ \cline{4-5}
& & & $h> h_{e, 2}$ & $K_3 \to E_2$ \\ \cline{1-1} \cline{4-5}
\multirow{2}{*}{$\Lambda > 0$, $\alpha\Lambda > 1/2$} &  &  & $h_2 > h > h_1$ & $nS \to nS$\\ \cline{4-5}
& & & $h>h_2$ & $K_3 \to nS$\\
      \hline
  \end{tabular}
\end{center}
\end{table}

So we summarize the regimes in Tables~\ref{D.3a} and~\ref{D.3b} . Apart from what we just mentioned about the regimes presented and regimes skipped, we also skip all ranges for $h$ which lies outside the domain of definition for $H(h)$ curves in Fig.~\ref{D3_1}.
In Table~\ref{D.3a} we presented regimes for $\alpha < 0$. One can see that for both $\Lambda < 0$ and $\Lambda >0$, $\alpha\Lambda \geqslant -5/8$
there are only two regimes -- $K_3 \to K_3$, and $K_3 \to E_{iso}$, and none of them is viable. For $\alpha\Lambda < -5/8$ we have two anisotropic exponential regimes, but both of them at large $\alpha\Lambda$ have
$H>0$ and $h>0$, so both
three-dimensional and extra-dimensional parts are expanding, which could violate the observations, so we cannot call them viable (at least, they are definitely less viable then those with contracting
extra dimensions). But at $\alpha\Lambda = -3/2$ the situation changes -- for $E_1$ we have $h=0$ with $H>0$ while for $E_2$ we have $H=0$ with $h>0$. With further decrease of $\alpha\Lambda < -3/2$ we have $h<0$ with $H>0$
for $E_1$ and $h>0$ with $H<0$ for $E_2$. Since in $D=3$ case both spaces are three-dimensional, it is unimportant which one is expanding while the other is contracting -- we call expanding one as ``our Universe'' while
contracting as extra dimensions, so for $\alpha\Lambda < -3/2$ we have two viable regimes. Let us finally note that this case is similar to $D=2$~\cite{my16b}.

For $\alpha > 0$ (see Table~\ref{D.3b}) we also report viable regimes. There we two of them: $K_3 \to E_1$ and $K_3 \to E_2$ for $\Lambda < 0$, the same two regimes for $\alpha > 0$, $\alpha\Lambda < 1/2$
and only one $K_3 \to E_2$ for $\alpha\Lambda = 1/2$. For $\alpha\Lambda > 1/2$ there are no viable regimes anymore.

\section{General $D\geqslant 4$ case}

In this case we use the general equations (\ref{H_gen})--(\ref{con2_gen}). The procedure is exactly the same as in the previous section -- we solve the constraint equation (\ref{con2_gen}) with respect to $H$ and obtain
three branches $H_1$, $H_2$, and $H_3$, then solve dynamical equations (\ref{H_gen})--(\ref{h_gen}) with respect to $\dot h$ and $\dot H$ and substitute individual branches to get $\dot h_i$ and $\dot H_i$ for each branch.

The expressions for $H_i(h)$ are lengthy so we do not write them down, but provide the results for the analysis. If we analyze the discriminant of (\ref{con2_gen}) with respect to $H$ and then the discriminant of the
resulting equation with respect to $h^2$ (exactly as it is done in the previous section), we obtain the critical values for $\zeta = \alpha\Lambda$:

\begin{equation}
\begin{array}{l}
\zeta_1 = - \dac{(D+2)(D+3)}{4D(D+1)},~\zeta_2 = \dac{\sqrt[3]{\mathcal{D}_2 (D-1)^2}}{12(D-2)(D-1)D(D+1)} + \\ \\ + \dac{(D^6 - 6D^5 + 10D^4 - 20D^2 + 24D + 36)(D-1)}{3D(D-2)(D+1)\sqrt[3]{\mathcal{D}_2 (D-1)^2}} +
\dac{D^3 - 9D^2 + 8D + 24}{12D(D-2)(D+1)},~\mbox{where} \\ \\
\mathcal{D}_2 = 10D^{10} + 6D^9 \mathcal{D}_1 - 100D^9 - 30D^8 \mathcal{D}_1 + 330 D^8 + 30D^7 \mathcal{D}_1 - 240 D^7 + 54D^6 \mathcal{D}_1 - \\ - 600D^6 - 84D^5 \mathcal{D}_1 + 240D^5 - 24D^4 \mathcal{D}_1 + 1520D^4 +
48D^3 \mathcal{D}_1 + 640 D^3 - 2880D^2 + 1728~\mbox{and}~ \\ \\ \mathcal{D}_1 = \dac{(D-4)(D-3)(D+2)}{(D-1)(D+1)}\sqrt{\dac{(D-4)(D+2)}{D(D-2)}}.
\end{array} \label{D.4_sep1}
\end{equation}

Additionally, for $\alpha < 0$, $\Lambda > 0$ there are two domains -- in one of them the resulting curves intersect $H=0$ while in the other they do not and these domains  are separated by

\begin{equation}
\begin{array}{l}
\zeta_3 = - \dac{D(D-1)}{4(D-2)(D-3)}.
\end{array} \label{D.4_sep2}
\end{equation}

\begin{figure}
\includegraphics[width=0.8\textwidth, angle=0]{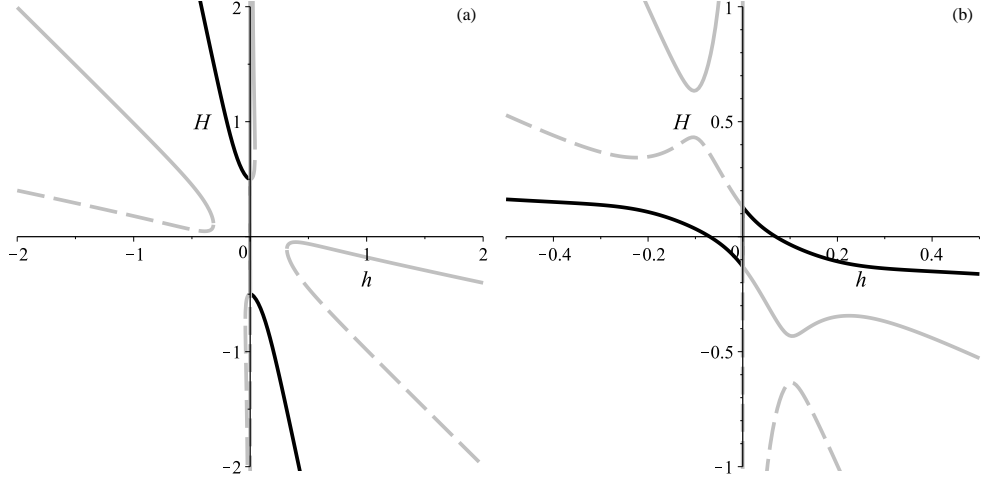}
\caption{Additional $H(h)$ graphs for $D\geqslant 4$ case: $\alpha < 0$, $\Lambda > 0$, $\alpha\Lambda < \zeta_3$ in (a) panel and $\alpha > 0$, $\Lambda > 0$, $\alpha\Lambda < \zeta_2$ in (b) panel.
Different branches presented by different linestyle/color combination: $H_1$ by solid black line, $H_2$ by solid grey and $H_3$ by dashed grey
(see the text for more details).}\label{D4_1}
\end{figure}

The resulting $H_i(h)$ curves in most of the cases resemble those from $D=3$ case. This happening for entire $\alpha < 0$ domain with different $\zeta_1$ for $D=3$ (where $\zeta_1 = -5/8$) and $D\geqslant 4$ cases, so the actual $H_i(h)$ curves for this case are presented in Figs.~\ref{D3_1}(a)--(c). The only difference between $D=3$ and $D\geqslant 4$ cases is that for the latter there is an additional regime presented in Fig.~\ref{D4_1}(a) for
$\alpha\Lambda < \zeta_3$. One can see that in the $D=3$ case the $H_i(h)$ curves always cross $H=0$ axis while in the general $D\geqslant 4$ case it is no longer the case 
(compare Fig.~\ref{D3_1}(c) with Fig.~\ref{D4_1}(a)). 
But, as we will state later, this does not affect
actual regimes -- the different is that the corresponding exponential solution ``moves'' from ($H>0$, $h>0$) to ($H<0$, $h>0$), but since both of them are non-realistic, this additional case with $\alpha\Lambda < \zeta_3$ does not affect the results.

For $\alpha > 0$ there are differences between these two cases -- the value for $\zeta_2$, which separate the regimes, is negative for $D=3$ and positive for $D\geqslant 4$, so
the differences also ``moved'' from $\Lambda < 0$ in $D=3$ to $\Lambda > 0$ in $D\geqslant 4$.
So the general $D\geqslant 4$ case with $\alpha > 0$, $\Lambda < 0$ looks exactly like $D=3$ $\alpha > 0$, $\Lambda < 0$ with $\alpha\Lambda > -1/8$, presented in Fig.~\ref{D3_1}(d); the situation presented in Fig.~\ref{D3_1}(e) does not appear in the general $D\geqslant 4$ case. Finally, due to the above mentioned shift of $\zeta_2$ into positive $\Lambda$ domain, we additionally have regime for $\alpha > 0$, $\Lambda > 0$, $\alpha\Lambda < \zeta_2$,
presented in Fig.~\ref{D4_1}(b); for $\alpha\Lambda \geqslant \zeta_2$ the situation looks exactly like in Fig.~\ref{D3_1}(f).

Let us also note that $\zeta_2$ is growing function of $D$ and

\begin{equation}
\begin{array}{l}
\lim\limits_{D\to\infty} \zeta_2 = \dac{\sqrt[3]{16}}{12} + \dac{\sqrt[3]{256}}{48} + \dac{1}{12} \approx 0.4256.
\end{array} \label{D.4_iso_1}
\end{equation}

One can see that on (a), (e), and (f) panels of Fig.~\ref{D3_1}
we have one isotropic solution ($H=h$), on (c) -- two and on (b) and (d) -- no isotropic solutions. There is a theoretical explanation: isotropic exponential solutions are governed by the equation

\begin{equation}
\begin{array}{l}
D(D+1)(D+2)(D+3)\alpha H^4 + (D+2)(D+3) H^2 - \Lambda = 0,
\end{array} \label{D.4_iso_1}
\end{equation}

\noindent and one can see that it is a biquadratic equation with respect to $H$. So that for solutions to exist we need not only positivity of the discriminant, but positivity of the roots. Then, skipping the derivation,
we can see that for $\alpha > 0$, $\Lambda < 0$ there are no isotropic solutions, for $\alpha > 0$, $\Lambda > 0$ as well as for $\alpha < 0$, $\Lambda < 0$ there is one, and for $\alpha < 0$, $\Lambda > 0$,
$\alpha\Lambda < \zeta_1$ there are no solutions while for $\alpha < 0$, $\Lambda > 0$, $\alpha\Lambda > \zeta_1$ there are two. In this regard, the scheme is quite the same as in $D=3$ case.

The existence of anisotropic exponential solutions is governed by the following equation

\begin{equation}
\begin{array}{l}
D^2 (D+1) (D-1)^2 (D-2) (4D^2 + 60D - 72)\xi^4 - D (D-1)^2 (40D^3 + 40D^2 + 96D - 288) \xi^3 + \\ + \[ D(D-1)(136 D^2 - 200D + 48)\zeta + (D-1) (16D^3 + 128D^2 - 24D - 144)  \] \xi^2 - \\
- \[ (104 D^2 - 152D + 48)\zeta + 12(D-2)(D-3)\]\xi + (4\zeta^2 + 12\zeta + 9) = 0,
\end{array} \label{D.4_aniso_1}
\end{equation}

\noindent where, as usual, $\xi = \alpha h^2$ and $\zeta = \alpha\Lambda$. If we consider and solve the discriminant of (\ref{D.4_aniso_1}) (we are not writing it down for it is 19th order polynomial in $D$), the
solutions are:

\begin{equation}
\begin{array}{l}
\zeta_4 = \dac{1}{2} \dac{D^2 - 4D + 6}{D(D-2)},~ \zeta_1,~\zeta_5 = \dac{3}{4} \dac{D(D+3)}{D^2 + 15D - 18},~ \zeta_6 = \dac{1}{4} \dac{3D^2 - 7D + 6}{D(D-1)},
\end{array} \label{D.4_aniso_2}
\end{equation}

\noindent where $\zeta_6$ is triple root and $\zeta_1$ is the same as in (\ref{D.4_sep1}). One can note that for $D<6$ $\zeta_4 > \zeta_5$ while for $D>6$ $\zeta_4 < \zeta_5$ and for $D=6$ they coincide.
So that $D=6$ is special case and its dynamics could be a bit
different -- we comment on it further.  Finally let us find $D\to\infty$ limits: $\lim\limits_{D\to\infty} \zeta_4 = 1/2$,
$\lim\limits_{D\to\infty} \zeta_1 = -1/4$, $\lim\limits_{D\to\infty} \zeta_5 = 3/4$, $\lim\limits_{D\to\infty} \zeta_6 = 3/4$, but $\zeta_6 > \zeta_5$ always.

The expressions for $\dot h(h)$ and $\dot H(h)$ are also lengthy so we will not write them down. When describing $H_i$ curves we mentioned that most of the cases coincide with those in $D=3$ case; the same is
true for the regimes. The $\alpha < 0$, $\Lambda < 0$ regimes are exactly the same as in $D=3$ case. The next, $\alpha < 0$, $\Lambda > 0$ regimes at
$\zeta \equiv \alpha\Lambda \geqslant \zeta_1$ are the same as in $D=3$ $\alpha\Lambda \geqslant -5/8$ case, and at $\zeta_1 < \zeta$ the regimes are the same as in $D=3$, $\alpha\Lambda < -5/8$ case. One
can see that $H(h)$ curves differ starting from $\zeta < \zeta_3$, but this does not change the regimes -- indeed, for both subcases we have the same anisotropic exponential solution $E_2$, but for $\zeta > \zeta_3$
it has $H > 0$ while for $\zeta < \zeta_3$ it has $H < 0$. But since this solution has $h > H$, it is not viable for both these cases. So that despite the differences in $H(h)$ curves, the regimes
for $\alpha < 0$ are the same as in $D=3$ case and the only viable one is the $K_3 \to E_1$ transition if $\alpha\Lambda \leqslant -3/2$ is fulfilled.

Now let us turn our attention to $\alpha > 0$ regimes. For $\Lambda < 0$ the $H(h)$ curves and the regimes are the same as in $D=3$ case with $\alpha\Lambda < -1/8$. The $\alpha\Lambda > -1/8$ counterpart from
$D=3$ does not exist in the general $D \geqslant 4$ case, as the separation $\zeta \equiv \alpha\Lambda$ value ``moved'' from negative values ($-1/8$ for $D=3$) to positive $\zeta_2$ from (\ref{D.4_sep1}). And since
the regimes are the same, the viable regime $K_3 \to E_1$ is also present in $D \geqslant 4$ case for the entire $\Lambda < 0$ range.
Finally,
$\alpha > 0$, $\Lambda > 0$ case exhibits the most interesting dynamics, similar to $D=3$ case. First of all, due to the differences in $H(h)$ structure, for $\zeta < \zeta_2$ (see Fig.~\ref{D4_1}(a)) the $E_1$ anisotropic
exponential solution has two $K_3$ regimes which lead to it, making the entire $h < 0$ range leads to viable compactification in $E_1$. For $\zeta \geqslant \zeta_2$ the situation is the same as in $D=3$ case
-- only $h < h_{e1}$ leads to $K_3 \to E_1$ transition. The second difference is in the ``fine structure'' of the additional exponential solutions. First it was described in $D=2$ and they appear in the
$15/32 \leqslant \zeta \leqslant 1/2$ range~\cite{my16b}, for $D=3$ they appear in the $3/8 \leqslant \zeta \leqslant 1/2$ range and for $D \geqslant 4$ case they appear for
$\min\{\zeta_4, \zeta_5\} \leqslant \zeta \leqslant \zeta_6$. By now we skip the description of the fine structure of these solutions and address it in the Discussions section. The realistic regimes in the general
case are the same as in $D=3$ -- it is $K_3 \to E_1$ from $H_2$ branch in the $H>0$, $h<0$ quadrant, and it manifest itself if $\alpha\Lambda \leqslant \zeta_6$.

So that the dynamics of the general $D \geqslant 4$ case is very similar to the $D=3$ one and the regimes are the same. The only differences are in details of the solutions (like in $\alpha < 0$, $\Lambda > 0$ case)
and the ranges, with the formers affect only non-realistic regimes. We can say that the regimes are the same as in $D=3$ case and could be seen in Tables~\ref{D.3a} and \ref{D.3b}. The realistic compactification
regimes exist for $\alpha < 0$, $\Lambda > 0$ with $\alpha\Lambda \leqslant 1/2$ and $\alpha > 0$ with $\alpha\Lambda \leqslant \zeta_6$, including entire $\Lambda < 0$.

\section{Discussions}

In this paper we investigated the existence and abundance of different regimes in $D=3$ and general $D \geqslant 4$ cases with $\Lambda$-term. Particular interest is paid to the solution which allow dynamical compactification.
Similar to the previously considered low-dimensional ($D=1, 2$) $\Lambda$-term cases~\cite{my16b}, the only viable dynamics is the transition from high-energy Kasner regime to anisotropic exponential expansion with expanding
three-dimensional space (``our Universe'') and contracting extra dimensional space. The majority of the non-viable regimes have nonstandard singularity as either future or past asymptote, and it is defined as follows.
As we can see from the equations of motion (\ref{dyn_gen}), they are nonlinear with
respect to the highest derivative\footnote{Actually, this is one of the definitions of Lovelock (and Gauss-Bonnet as its particular case) gravity: it is well-known~\cite{etensor1, etensor2, etensor3} that
the Einstein tensor is, in any dimension, the only symmetric and
conserved tensor depending only on the metric and its first and
second derivatives (with a linear dependence on second
derivatives). If one drops the condition of linear dependence on
second derivatives, one can obtain the most general tensor which
satisfies other mentioned conditions -- Lovelock
tensor~\cite{Lovelock}.}, so formally we can solve them with respect to it. Then, the highest derivative is expressed as a ratio of two polynomials, both
depending on $H$. And there could be a situation when the denominator of this expression is equal to zero while the numerator is not. In this case $\dot H$ diverges while $H$ is (generally) nonzero and
regular. In our study we saw nonstandard singularities with divergent $\dot h$ or both $\dot h$ and $\dot H$ at nonzero or sometimes zeroth $H$.
This kind of singularity is ``weak'' by Tipler's classification~\cite{Tipler}, and ``type II''
in classification by Kitaura and Wheeler~\cite{KW1, KW2}. Recent studies of the singularities of this kind in the cosmological context in Lovelock and Einstein-Gauss-Bonnet gravity
demonstrate~\cite{CGP2, mpla09, grg10, KPT, prd10} that their presence is not suppressed and they are abundant for a wide range of initial conditions and parameters and sometimes~\cite{prd10} they are the
only option for future behavior.

Below we summarize our findings for $D=3$ and general $D \geqslant 4$ $\Lambda$-term cases and discuss the general results for the
$\Lambda$-term case (with~\cite{my16b} taken into account).

First case to consider, $D=3$, demonstrate two viable regimes $K_3 \to E_{1, 2}$ for $\alpha < 0$, $\Lambda > 0$ and $\alpha\Lambda \leqslant -3/2$ (see Table \ref{D.3a}) and two more regimes $K_3 \to E_{1, 2}$ for
$\alpha > 0$, $\alpha\Lambda \leqslant 1/2$, including $\Lambda < 0$ (see Table \ref{D.3b}). The former of them in $\alpha\Lambda = -3/2$ limiting case have either $h=0$ (for $E_1$) or $H=0$ (for $E_2$); the latter
 in the limiting case $\alpha\Lambda = 1/2$ have only $K_3 \to E_2$ transition.

The second, general $D \geqslant 4$ case, have dynamics very similar to $D=3$ case and so the regimes -- for $\alpha < 0$, $\Lambda > 0$, $\alpha\Lambda \leqslant -3/2$ we have $K_3 \to E_1$ transition and again, similar to the
$D=2, 3$ cases, have $h=0$ for $\alpha\Lambda = -3/2$. Another viable regime for the general case takes place at $\alpha > 0$, $\alpha\Lambda \leqslant \zeta_6$ from (\ref{D.4_aniso_2}) (including $\Lambda < 0$).
So the regimes are exactly the same, but the coverage of the second regime -- with $\alpha > 0$ -- is different -- in $D=3$ we have this regime if $\alpha\Lambda \leqslant 1/2$ while in the general $D$ cases it is
$\alpha\Lambda \leqslant \zeta_6$, with greater area on $(\alpha, \Lambda)$ space than in $D=3$ case.

To summarize all $\Lambda$-term cases, the only pathological one is $D=1$ -- all the remaining have viable $K_3 \to E_{3+D}$ transition over an open range of parameters: for $D=2$ we have viable solutions in two
domains -- $\alpha > 0$, $\alpha\Lambda \leqslant 1/2$ and $\alpha < 0$, $\Lambda > 0$, $\alpha\Lambda \leqslant -3/2$. In $D=3$ we have ``doubled'' number of regimes in the same two domains. 
We used the term ``doubled'' because in $D=3$ both spaces are
three-dimensional and so it is irrelevant which one is expanding and which one is contracting -- the expanding one is ``our Universe'' and the contracting one is extra-dimensional.
Finally, in the general $D \geqslant 4$ case we have the same regimes but with greater coverage over $(\alpha, \Lambda)$ plane. So that we can conclude that with increase of the number of extra dimensions, the
occurrence of the transition from high-energy Kasner to anisotropic exponential solution $K_3 \to E_{3+D}$ (realistic compactification) is increasing.

This way the viable compactification regimes are the same in all $D \geqslant 2$ cases (but with different coverage), but the fine structure of other exponential solutions in $\alpha > 0$, $\Lambda > 0$ domain
is different. We described it in detail in
$D=2$ case (see~\cite{my16b}), in less detail in $D=3$ and skipped for the general $D \geqslant 4$ case, so now let us briefly summarize this structure for all different cases. Of additional interest is
mentioned earlier $D=6$ case -- with $\zeta_4 = \zeta_5$ it demonstrate a bit different behavior. So we summarized the regimes for $H_2-H_3$ branches for viable values for $H$ and $h$ ($h<0$, $H>0$) in
Table~\ref{D.4b}.

\begin{table}
\begin{center}
\caption{Fine structure of $\alpha > 0$ $\Lambda > 0$ exponential solutions.}
\label{D.4b}
  \begin{tabular}{|c|c|c|}
    \hline
   $D$ & $\alpha\Lambda$ &  Regimes   \\
    \hline
\multirow{7}{*}{$D=2$} & $\alpha\Lambda < 15/32$ & $K_3 \to E_1 \leftarrow nS \to nS \leftarrow nS^{(-)} $ \\ \cline{2-3}
& $\alpha\Lambda = 15/32$ & $K_3 \to E_1 \leftarrow nS \to E_{2, 3} \to nS \leftarrow nS^{(-)} $ \\ \cline{2-3}
& $1/2 > \alpha\Lambda > 15/32$ & $K_3 \to E_1 \leftarrow nS \to E_3 \leftarrow E_2 \to nS \leftarrow nS^{(-)} $ \\ \cline{2-3}
& $\alpha\Lambda = 1/2$ & $K_3 \to E_1 \leftarrow E_{2, 3} \to nS \leftarrow nS^{(-)}$ \\ \cline{2-3}
& $1> \alpha\Lambda > 1/2$ & $K_3 \to nS \leftarrow E_1 \to nS \leftarrow nS^{(-)}$ \\ \cline{2-3}
& $\alpha\Lambda = 1$ & $K_3 \to nS \leftarrow E_1 \to nS$ \\ \cline{2-3}
& $\alpha\Lambda > 1$ & $K_3 \to nS \leftarrow E_1 \to E_1^{(-)}$ \\ \cline{1-3}
\multirow{5}{*}{$D=3$} & $\alpha\Lambda < 3/8$ & $K_3 \leftarrow E_2 \to nS \leftarrow nS \to E_1 \leftarrow K_3$ \\ \cline{2-3}
&  $\alpha\Lambda = 3/8$ & $K_3 \leftarrow E_2 \to nS \leftarrow E_{3, 4} \leftarrow nS \to E_1 \leftarrow K_3$ \\ \cline{2-3}
& $1/2 > \alpha\Lambda > 3/8$ & $K_3 \leftarrow E_2 \to nS \leftarrow E_{4} \to E_3 \leftarrow nS \to E_1 \leftarrow K_3$ \\ \cline{2-3}
& $\alpha\Lambda = 1/2$ & $K_3 \leftarrow E_2 \to E_1 \leftarrow K_3$ \\ \cline{2-3}
& $\alpha\Lambda > 1/2$ & $K_3 \leftarrow nS \to nS \leftarrow K_3$ \\ \cline{1-3}
\multirow{8}{*}{$D=4, 5 \cup D > 6$} & $\alpha\Lambda < \zeta_2$ & $K_3 \to E_1 \leftarrow K_3$ \\ \cline{2-3}
& $\zeta_5 > \alpha\Lambda \geqslant \zeta_2$ & $K_3 \leftarrow E_2 \to nS \leftarrow nS \to E_1 \leftarrow K_3$ \\ \cline{2-3}
& $\alpha\Lambda = \zeta_5$ & $K_3 \leftarrow E_2 \to nS \leftarrow E_{3, 4} \leftarrow nS \to E_1 \leftarrow K_3$ \\ \cline{2-3}
& $\zeta_4 > \alpha\Lambda > \zeta_5$ & $K_3 \leftarrow E_2 \to nS \leftarrow E_4 \to E_3 \leftarrow nS \to E_1 \leftarrow K_3$ \\ \cline{2-3}
& $\alpha\Lambda = \zeta_4$ & $K_3  \leftarrow E_4 \to E_3 \leftarrow nS \to E_1 \leftarrow K_3$ \\ \cline{2-3}
& $\zeta_6 > \alpha\Lambda > \zeta_4$ & $K_3  \leftarrow nS \to E_3 \leftarrow nS \to E_1 \leftarrow K_3$ \\ \cline{2-3}
& $\alpha\Lambda = \zeta_6$ & $K_3 \leftarrow nS \to E_1 \leftarrow K_3$ \\ \cline{2-3}
& $\alpha\Lambda > \zeta_6$ & $K_3 \leftarrow nS \to nS \leftarrow K_3$ \\ \cline{1-3}
\multirow{5}{*}{$D=6$} & $\zeta_{4, 5} > \alpha\Lambda \geqslant \zeta_2$ & $K_3 \leftarrow E_2 \to nS \leftarrow nS \to E_1 \leftarrow K_3$ \\ \cline{2-3}
& $\alpha\Lambda = \zeta_{4, 5}$ & $K_3 \leftarrow E_2 \leftarrow nS \to E_1 \leftarrow K_3$ \\ \cline{2-3}
& $\zeta_6 > \alpha\Lambda > \zeta_{4, 5}$ & $K_3 \leftarrow nS \to E_3 \leftarrow nS \to E_1 \leftarrow K_3$ \\ \cline{2-3}
& $\alpha\Lambda = \zeta_6$ & $K_3 \leftarrow nS \to E_1 \leftarrow K_3$ \\ \cline{2-3}
& $\alpha\Lambda > \zeta_6$ & $K_3 \leftarrow nS \to nS \leftarrow K_3$ \\
\hline
  \end{tabular}
\end{center}
\end{table}

From Table~\ref{D.4b} one can clearly see realistic compactification regimes and restrictions on $\alpha\Lambda$ when they occur. In $D=2$ it is $K_3 \to E_1$ at the beginning of the regimes string and one
can clearly see that it occurs only for $\alpha\Lambda \leqslant 1/2$ (the arrow indicate the regime transition with respect to the ``standard'' time direction -- from past asymptote to future asymptote). 
In $D=3$ it is also $K_3 \to E_1$ but now it is in the end of regimes string and it also occurs only for $\alpha\Lambda \leqslant 1/2$.
Similarly, for general $D \geqslant 4$ (including $D=6$) it is $K_3 \to E_1$ in the end of regimes string and it occurs for $\alpha\Lambda \leqslant \zeta_6$.  So that one can see that the fine structure of the
regimes is different in different $D$ but it does not affect the realistic regimes. One last note, as we mentioned above, for $D<6$ $\zeta_4 > \zeta_5$ while for $D>6$ $\zeta_4 < \zeta_5$, so that for $D > 6$
$\zeta_4$ and $\zeta_5$ should be exchanged (in Table~\ref{D.4b} they input as $\zeta_4 > \zeta_5$).

At this point it is appropriate to address another point -- in~\cite{my16b} we mentioned that some of the exponential solutions have directional stability. It is also could be clearly seen from Table~\ref{D.4b} --
the solutions ($\to E \leftarrow$) are stable, the solutions ($\leftarrow E \to$) are unstable and finally the solutions ($\to E \to$) have directional stability. The stability of exponential solutions in EGB and Lovelock
gravity was addressed in~\cite{my15} and for general EGB case in~\cite{iv16}. The results of these studies could be summarized as follows -- the exponential solution with nonzero total expansion rate ($\sum H_i$) is
stable if $\sum H_i > 0$ and unstable if $\sum H_i < 0$. All solutions -- both stable and unstable -- follow this rule, but in this scheme there is no place for directional stability, it could seem. But in
reality there is -- indeed, the solutions with directional stability have $\sum H_i = 0$ -- constant volume solutions (see~\cite{CST2} for more detail). So that in the direction which give $\sum H_i > 0$ they are
stable while in the opposite direction -- with $\sum H_i < 0$ -- unstable. This situations is very well illustrated in Figs.\ref{D3_4} (c, d) -- on (c) panel we
have $nS \to E \to nS$ and it demonstrate directional stability -- indeed, for $H > H_E$ we have $\sum H_i > 0$ while for $H < H_E$ it is $\sum H_i < 0$. On the
(d) panel the regimes are $nS \to E_1 \leftarrow E_2 \to nS$ and $E_1$ is stable (it has $\sum H_i > 0$) while $E_2$ is unstable (it has $\sum H_i < 0$).

\begin{figure}
\includegraphics[width=1.0\textwidth, angle=0]{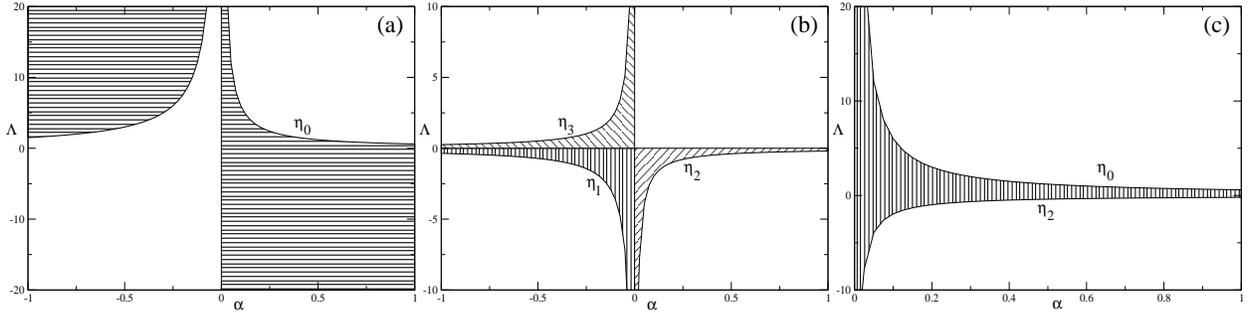}
\caption{Summary of the bounds on ($\alpha, \Lambda$) from this paper alone on (a) panel; from other considerations found in the literature on (b) panel and the intersection between them
on (c) panel
(see the text for more details).}\label{DD}
\end{figure}

As we have found the domains on ($\alpha$, $\Lambda$) plane which give us realistic regimes, it is interesting to compare these bounds with those coming from other considerations. The results
of this comparison are presented in Fig.~\ref{DD}. In Fig.\ref{DD}(a) we presented the summary of the results from the current paper -- $\alpha < 0$, $\Lambda > 0$, 
$\alpha\Lambda \leqslant -3/2$ in the second quadrant and $\alpha > 0$, $\alpha\Lambda \leqslant \eta_0 \equiv \zeta_6$ from (\ref{D.4_aniso_2}) on $\alpha > 0$ half-plane. In Fig.\ref{DD}(b)
we collected all available constraints on $\alpha\Lambda$ from other considerations. Among them a significant part is based on the different aspects of Gauss-Bonnet gravity in AdS spaces --
from consideration of shear viscosity to entropy ratio as well as causality violations and CFTs in dual gravity description there were obtained limits
on $\alpha\Lambda$~\cite{alpha_01, alpha_02, alpha_03, alpha_04, alpha_05, alpha_06, alpha_07, alpha_08}:

\begin{equation}
\begin{array}{l}
 - \dac{(D+2)(D+3)(D^2 + 5D + 12)}{8(D^2 + 3D + 6)^2} \equiv \eta_2 \leqslant \alpha\Lambda \leqslant \eta_1 \equiv \dac{(D+2)(D+3)(3D + 11)}{8D(D+5)^2}.
\end{array} \label{alpha_limit}
\end{equation}

The limits for dS ($\Lambda > 0$) are less numerous and are based on different aspects (causality violations, perturbation propagation and so on) of black hole physics in dS spaces. The most
stringent constraint coming from these considerations is~\cite{add_rec_2, add_rec_4, dS}

\begin{equation}
\begin{array}{l}
\alpha\Lambda \geqslant \eta_3 \equiv - \dac{D^2 + 7D + 4}{8(D-1)(D+2)}.
\end{array} \label{alpha_limit2}
\end{equation}

At this point, two clarifications are required. First, this limit is true for both $\alpha\lessgtr 0$ and $\Lambda \lessgtr 0$, so that in the $\alpha > 0$, $\Lambda < 0$ quadrant two limits
are applied: $\alpha\Lambda \geqslant \eta_2$ from (\ref{alpha_limit}) and $\alpha\Lambda \geqslant \eta_3$ from (\ref{alpha_limit2}). One can easily check that $\eta_2 > \eta_3$ for 
$D \geqslant 2$ so that the constraint from (\ref{alpha_limit}) is the most stringent in this quadrant. Secondly, one can see that the limit in (\ref{alpha_limit2}) is not defined for
$D=1$. Indeed, in this case the limit is special (see~\cite{alpha_12}), but for $D=1$ there are no viable cosmological regimes (see~\cite{my16b}), so that we consider $D \geqslant 2$ only.

One can see that the bounds on ($\alpha, \Lambda$) cover three quadrants and our analysis allows to constraint the remaining sector: $\alpha > 0$, $\Lambda > 0$. But if we consider joint
constraint from both Figs.\ref{DD}(a) and (b), the resulting area is presented in Fig.\ref{DD}(c). In there, one can see that the regimes in $\alpha < 0$ sector disappear due to the fact that
$\eta_3 > -3/2$ always---the ($\alpha, \Lambda$) which have viable cosmological dynamics in $\alpha < 0$ sector disagree with (\ref{alpha_limit2}). To conclude, if we consider our bounds on
($\alpha$, $\Lambda$) together with previously obtained (see (\ref{alpha_limit})--(\ref{alpha_limit2})), the resulting bounds are

\begin{equation}
\begin{array}{l}
\alpha > 0, \quad D \geqslant 2, \quad \dac{3D^2 - 7D + 6}{4D(D-1)}  \equiv \eta_0 \geqslant \alpha\Lambda \geqslant \eta_2 \equiv - \dac{(D+2)(D+3)(D^2 + 5D + 12)}{8(D^2 + 3D + 6)^2}.
\end{array} \label{alpha_limit3}
\end{equation}

The result that the joint analysis suggests only $\alpha > 0$ is interesting and important -- indeed, the constraints on ($\alpha, \Lambda$) considered so far do not distinguish between
$\alpha \lessgtr 0$, and there are several considerations which favor $\alpha > 0$. The most important of them is the positivity of $\alpha$ coming from heterotic string setup, where $\alpha$
is associated with inverse string tension~\cite{alpha_12}, but there are several others like ill-definition of the holographic entanglement entropy~\cite{entr}. So that, our joint analysis
support $\alpha > 0$ as well.

\section{Conclusions}

In this paper we finalized study of different regimes in EGB cosmologies with $\Lambda$-term. We compared the regimes existence and abundance between different $D$ within $\Lambda$-term cases, and now it is time to
compare between $\Lambda$-term and vacuum cases.

The main difference between $\Lambda$-term and vacuum cases is the absence of ``Kasner transition'' (transitions from high-energy (Gauss-Bonnet) Kasner to low-energy (GR) one) in the $\Lambda$-term case and its
presence in the vacuum. The reason for this difference is simple -- as we demonstrated in~\cite{my16b}, in the presence of $\Lambda$-term power-law solutions do not exist. Some sort of unstable $K_1$ we reported
in~\cite{my16b} for $D=1$ $\Lambda$-term case, but it is singular and is never reached so it is unphysical. Formally, high-energy Kasner regime $K_3$ also should not exist, but in the high-energy limit
$H_i \ggg \Lambda$ and so $\Lambda/H_i \lll 1$ and we can treat high-energy $\Lambda$-term regime as vacuum. So that for vacuum cases we have as viable both ``Kasner transitions'' and transitions from GB Kasner
to anisotropic exponential solutions, while for $\Lambda$-term cases we have only the latter. But for vacuum cases we have only one viable exponential regime while for $\Lambda$-term we have two. Also, for vacuum cases
viable exponential regimes exist only for $\alpha > 0$ while for $\Lambda$-term one of them exists for $\alpha > 0$ while another for $\alpha < 0$.

This is another difference between vacuum and $\Lambda$-term solutions -- the abundance of the exponential solutions. This topic for anisotropic was investigated in~\cite{CPT3} and one can clearly see that the number
of the solutions is substantially decrease in the vacuum case. The case of isotropic solutions is more easier to address -- indeed, in the isotropic case (\ref{H_gen})--(\ref{con2_gen}) reduce to a single equation

\begin{equation}
\begin{array}{l}
\alpha D (D+1)(D+2)(D+3) H^4 + (D+2)(D+3)H^2 = \Lambda.
\end{array} \label{concl_1}
\end{equation}

In the vacuum case $\Lambda \equiv 0$ (\ref{concl_1}) has only one root $H^2 = - 1/(\alpha D(D+1))$ while in the $\Lambda$-term case we could have up to two roots. And this is exactly what we observe -- in the vacuum
case~\cite{my16a} we always have only one isotropic solution while in $\Lambda$-term cases (\cite{my16b} and this paper) we have up to two of them. Similarly, if we consider higher (say, $n$th order) Lovelock orders,
we shall have up to $n$ isotropic solutions in the $\Lambda$-term case and up to $(n-1)$ in the vacuum.

The results of our paper are the list of all regimes and the bounds on ($\alpha$, $\Lambda$) where the viable cosmologies exist within. The latter could be compared with similar bounds
but from other considerations, and this comparison is presented in Fig.\ref{DD}; the joint analysis (Fig.\ref{DD}(c)) suggests that only $\alpha > 0$, $D \geqslant 2$ is allowed.

This finalize our paper and the discussion of its results. We claim that we thoroughly investigated the case under consideration and analytically obtained the bounds on ($\alpha$, $\Lambda$)
which allow the existence of the cosmologically viable regimes. Our bounds could be confronted with other similar bounds and the comparison could lead to interesting conclusions.
The investigation of the viable Gauss-Bonnet cosmologies does not end here---there are other interesting effects like curvature and other matter sources which could affect the viability,
and we shall consider these effects in the near future.

\begin{acknowledgments}
This work was supported by FAPEMA under project BPV-00038/16.
\end{acknowledgments}

\end{document}